\documentclass[12pt]{article}

\usepackage[numbers,sort&compress]{natbib}
\usepackage{times}
\usepackage{multibib}

\usepackage{color}
\usepackage{soul}

\usepackage{diagbox}
\usepackage[table]{xcolor} 
\usepackage{tikz,subcaption}
\usetikzlibrary{positioning}
\usepackage{array, multirow}
\newcolumntype{L}{>{\arraybackslash}m{2.375cm}}

\newcommand\T{\rule{0pt}{2.6ex}}       
\newcommand\B{\rule[-1.2ex]{0pt}{0pt}} 

\usepackage{caption}
\usepackage{fix-cm}

\usepackage{amsmath,amssymb, mathtools}

\usepackage{graphicx}

\usepackage[right]{lineno}

\newcommand{\ASYM}{{\textsc{\tiny ASYM}}}

\newcommand{\GEN}{{\textsc{\tiny GEN}}}
\newcommand{\TIPS}{{\textsc{\tiny TIPS}}}
\newcommand{\TOT}{{\textsc{\tiny TOT}}}
\newcommand{\TIP}{{\textsc{\tiny TIP}}}
\newcommand{\N}{{\textsc{\tiny N}}}

\newenvironment{sciabstract}{%
\begin{quote} \bf}
{\end{quote}}

\title{Title:  Branching principles of animal and plant networks identified by combining extensive data, machine learning, and modeling \\
Short Title:  Functional groups of biologic branching}

\author
{Authors:  Alexander Byers Brummer$^{1,2,3\ast}$, \\ 
Panagiotis Lymperopoulos$^{1}$, Jocelyn Shen$^{4}$, \\
Elif Tekin$^{1,2}$, Lisa P. Bentley$^{5}$, Vanessa Buzzard$^{6}$, \\
Andrew Gray$^{6}$, Imma Oliveras$^{7}$, \\
Brian J. Enquist$^{6, 8}$, Van M. Savage$^{1,2,3,8}$\\
\\
\normalsize{$^{1}$Institute for Quantitative and Computational Biology, University of California, Los Angeles}\\
\normalsize{$^{2}$Department of Computational Medicine, David Geffen School of MedicineUniversity of California, Los Angeles}\\
\normalsize{$^{3}$Department of Ecology and Evolutionary Biology, University of California, Los Angeles}\\
\normalsize{$^{4}$ Department of Electrical Engineering and Computer Science, Massachusetts Institute of Technology}\\
\normalsize{$^{5}$ Department of Biology, Sonoma State University}\\
\normalsize{$^{6}$ Department of Ecology and Evolutionary Biology, University of Arizona}\\
\normalsize{$^{7}$ Environmental Change Institute, University of Oxford}\\
\normalsize{$^{8}$ Santa Fe Institute}\\
\\
\normalsize{$^\ast$To whom correspondence should be addressed; E-mail:  abrummer@ucla.edu.}
}

\date{}

\begin{document} 

\baselineskip24pt


\maketitle 

\linenumbers


\begin{sciabstract}
One Sentence Summary:  Vascular theory and machine-learning enable principled classification of the form and function of biologic branching in mammals and plants.

Abstract:  Branching in vascular networks and in overall organismic form is one of the most common and ancient features of multicellular plants, fungi, and animals.  By combining machine-learning techniques with new theory that relates vascular form to metabolic function, we enable novel classification of diverse branching networks\textemdash mouse lung, human head and torso, angiosperm and gymnosperm plants.  We find that ratios of limb radii\textemdash which dictate essential biologic functions related to resource transport and supply\textemdash are best at distinguishing branching networks.  We also show how variation in vascular and branching geometry persists despite observing a convergent relationship across organisms for how metabolic rate depends on body mass.
  
\end{sciabstract}


\paragraph*{}

\section*{Introduction}

\begin{figure}
\begin{centering}
\includegraphics[width = 0.5\textwidth]{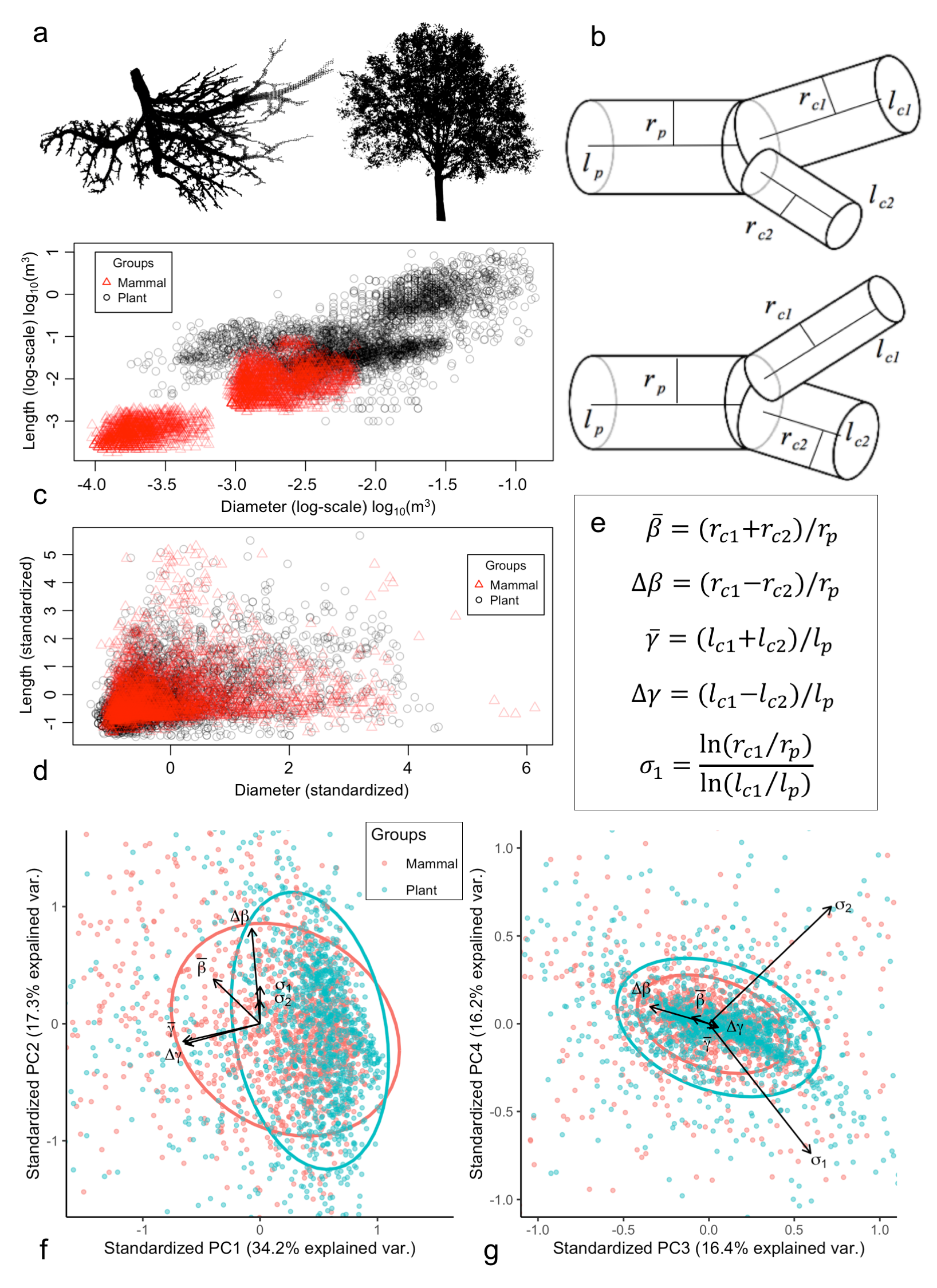}
\caption[Tree diagrams and machine learning results.]{\label{fig:tree_pca} \fontsize{10}{8}\selectfont  \textbf{a} Examples of mouse lung and angiosperm branching networks. \textbf{b} Diagrams of \textit{positive} (top) and \textit{negative} (bottom) asymmetric branching. \textbf{c} Scatter plot of lengths and diameters of all data studied, logarithmically scaled, shows trivial size-based clustering.  \textbf{d}  Scatter plot of standardized (zero mean and unit variance) lengths and diameters of all data studied shows non-informative overlap.  \textbf{e} Definitions of biophysically motivated transformations: average and difference radial scale factors ($\bar{\beta}, \Delta \beta$) related to hydraulic resistance, length scale factors ($\bar{\gamma}, \Delta \gamma$) related to space-filling, and sibling slenderness scaling exponents ($\sigma_1, \sigma_2$) related to gravitational bending and buckling.  \textbf{f} and \textbf{g} First through fourth principle components of variables defined in \textbf{e}, ellipses are contours of 75\% quantiles for bivariate principle components, and vector lengths indicate PC loadings.  \textbf{f} PC1 and PC2 show large extent of variance associated with radial and length scale factors, with group clustering determined separately by $\Delta \beta$ for plants and $\bar{\beta}$ for mammals (see Figure 3).  \textbf{g} PC3 and PC4 show variances due to asymmetric radial scaling ($\Delta \beta$) and linear combinations of sibling slenderness scaling exponents.}
\end{centering}
\end{figure}

It is a great challenge to decipher which features of biological branching networks are shared, which are different, and when these differences matter \cite{price_etal_ecolett_2012, alilou_etal_scireports_2018}.  For instance, branching in plant and animal networks exhibit strikingly similar features despite profound physiological and environmental differences (e.g., carbon dioxide and sap versus oxygen and blood, mobile versus stationary organisms, heart and pulsatile flow versus non-pulsatile flow) \cite{bentley_etal_ecolett_2013, conn_etal_cellsys_2017, newberry_etal_ploscompbio_2015, smith_etal_newphyt_2014, savage_etal_ploscompbio_2008, ronellenfitsch_etal_prl_2016, tekin_etal_ploscompbio_2016, west_etal_science_1997, west_etal_nature_1999}.  Similarly, differences in loopiness and ``noisiness" are well documented between vascular branching in tumors or stroke-damaged tissue versus healthy tissue \cite{alilou_etal_scireports_2018, jain_science_2005, prakash_etal_stroke_2013}.  The shared branching features are argued to lead to functional convergence in plant and animal networks via biological rates despite the notable physiological differences just listed \cite{savage_etal_funceco_2004, mori_etal_pnas_2010, kolokotrones_etal_nature_2010}.  Yet, the extent of shared versus distinct branching features has not been systematically and quantitatively analyzed across plants and animals in the same study.  Consequently, there is a need to understand the forces that shape the full spectrum of form and function in branching networks (Figure \ref{fig:tree_pca}\textbf{a}).

The classification of branching architectures is historically based on coarse qualitative differences in morphological features.  Examples include:  classifying lobes of the liver based on independent blood supply \cite{couinaud_liver_1957}; or the paired/un-paired ordering of plant leaves along a stem \cite{hofmeister_plant_growth_phyllotaxy_1868}.   Recent efforts have identified gene expression profiles related to branching phenotypes, with examples of branching in the developing lung as being planar versus tetrahedral in orientation \cite{metzger_etal_nature_2008}, or branching in the developing kidney based on the number of terminal vessels downstream from (or distal to) two sibling branches \cite{lefevre_etal_development_2017}.  However, these empirically motivated classifications still fall short of relating patterns in vascular form to biophysical and biomechanical function.

With recent advances in automated methods of image analysis developed by us and others  \cite{conn_etal_cellsys_2017, newberry_etal_ploscompbio_2015, lau_etal_trees_2018}, increasing amounts of data are becoming available to tackle these problems.  The tools that are missing are efficient and accurate algorithms for categorizing branching across whole networks and different organisms.  In this paper we apply machine-learning methods to theoretically-informed feature spaces to leverage all available information and technology to achieve these goals.

We analyze the largest-ever compilation of branching network data, with over 58 distinct networks and approximately 8,000 vessels or tree limbs (see Table \ref{tab:network_measures}).  We collected these data over the last decade for both mammalian cardiovascular systems and plant architecture in both angiosperms and gymnosperms.  Two mammalian networks are studied, the first being the major arterial branching junctions of the human head and torso (HHT) for 18 adult individuals (\textit{H. sapiens}) collected using contrast-enhanced magnetic resonance angiography on a 3 Tesla Siemens Trio scanner with voxel dimensions between $700 \times 700 \times 800 \mu m^3$ and $800 \times 800 \times 900 \mu m^3$ \cite{newberry_etal_ploscompbio_2015}.  The second mammalian network is the full pulmonary vascular branching of one wild-type adult mouse lung (ML) (\textit{M. musculus}) collected using a combination of vascular casting with MICROFIL and micro computed tomography on a $\mu$CT 40, ScanCo Medical scanner with 10$\mu$m isotropic voxel spacing \cite{tekin_etal_ploscompbio_2016}.  All mammalian network data was acquired using the open source software Angicart \cite{newberry_angicart}.

\begin{table}
\caption[Basic measures of the different vascular branching networks studied.]
{\label{tab:network_measures}\textbf{$|$ Basic measures of the different vascular branching networks studied.}}
\begin{centering}
\resizebox{\textwidth}{!}{%
\begin{tabular}{L|L|L|L|L|L|L|L|L}
        \hline
        \multicolumn{1}{m{3.25cm}|}{Vascular branching network} & \multicolumn{1}{m{2cm}|}{Trunk radius (cm)} & \multicolumn{1}{m{2cm}|}{Trunk length (cm)} & \multicolumn{1}{m{2.25cm}|}{Mean tip \newline radius (mm)} & \multicolumn{1}{m{2.25cm}|}{Mean tip \newline length (mm)} & \multicolumn{1}{m{1.5cm}|}{Number of tips} & \multicolumn{1}{m{2cm}|}{Number of generations} & \multicolumn{1}{m{2cm}|}{Number of junctions} & \multicolumn{1}{m{2.25cm}}{Total number of vessels} \\
        \hline 
        ML (N=1) & 0.0686 & 0.103 & 0.098 (0.055) & 0.709 (0.434) & 688 & 9 & 660 & 1348\\
        \hline
        HHT (N=18) & 0.383 (0.09) & 2.77 (1.74) & 0.855 (0.474) & 6.950 (7.18) & 50 (30) & 6 (1) & 48 (30) & 1891\\
        \hline
        Balsa (N=1) & 18.8 & 1170 & 5.97 (4.99) & 125 (151) & 357 & 8 & 292 & 649\\
        \hline
        Pi\~{n}on (N=1) & 5.73 & 5.4 & 2.30 (1.61) & 24.5 (18.2) & 1286 & 10 & 813 & 2099\\
        \hline
        \multicolumn{1}{m{3cm}|}{Ponderosa (N=5)} & 2.06 (0.769) & 30.6 (30.1) & 2.25 (0.714) & 77.8 (62.0) & 31 (21) & 5 (1) & 23 (21) & 312\\
        \hline
        \multicolumn{1}{m{3cm}|}{Roots (N=314)} & 0.307 (0.293) & 25.9 (23.8) & 1.19 (0.838) & 89.2 (74.1) & 2 (2) & 1 (1) & 1 (1) & 1231\\
        \hline
        \multicolumn{1}{m{3.5cm}|}{AS/GS Tips (N=31)} & 0.320 (0.103) & 10.8 (6.3) & 0.516 (0.318) & 57.8 (53.7) & 15 (11) & 4 (1) & 12 (9) & 914\\
        \hline
    \end{tabular}%
    }
\caption*{\\ \fontsize{10}{8}\selectfont Physical dimensions and counts of various network properties, including: initial (trunk) and terminal (tip) vessels and branches.  For single network datasets ($N = 1$) reported values are exact.  For multi-network datasets ($N > 1$), values are averages with standard deviations reported in parenthesis.  For a given network, the number of generations, $N_{\GEN}$, is determined from number of tips p, $N_{\TIPS}$, as $N_{\GEN} = \ln(N_{\TIPS})/\ln(2)$, and rounded to the nearest integer.  Due to approximate log-normality of distributions, means and standard deviations were determined in log-space and back transformed.  Total number of vessels }
\end{centering}
\end{table}

The plant networks consist of: 1. whole, above-ground, adult trees for one Balsa (\textit{O. pyramidale}), one Pi\~{n}on (\textit{P. edulis}), and five Ponderosa pines (\textit{P. ponderosa}) \cite{bentley_etal_ecolett_2013}, 2. an array of angiosperm root clusters belonging to Andean tropical montane cloud forests \cite{oliveras_etal_plantecodiv_2014}, and 3. a collection of 50 \textit{cm} long clippings of the terminal ends of canopy branches from three species each of angiosperms (AS Tips) and gymnosperms (GS Tips) comprised of Maple (\textit{A. grandidentatum}), Scrub Oak (\textit{Q. gambelii}), Robinia (\textit{R. neomexicana}), White Fir (\textit{A. concolor}), Douglas Fir (\textit{P. menziesii}), and White Pine (\textit{P. strobiformis}).  Tree measurements\textemdash all done destructively by hand\textemdash are of the external branching structures (limbs), not the xylem that are directly responsible for water transport.  Scaling relationships for the external limbs directly determine similar relationships for the internal xylem based on previous empirical studies \cite{olson_etal_ecolett_2014, reich_jeco_2014} and established branching theory \cite{west_etal_nature_1999, savage_etal_pnas_2010}, thus enabling comparisons of plant and animal networks for the structure, flow, and function in the present study \cite{bentley_etal_ecolett_2013, lau_etal_trees_2018, eloy_etal_natcomm_2017}.

To search for patterns, machine-learning is often applied to the full set of untransformed, standardized raw data.  These raw data thus represent one feature space, yet there are always infinitely more choices of feature spaces based on specific combinations, subsets, mathematical operations (e.g., logarithms or ratios), or other transformations of the raw data (Figure \ref{fig:tree_pca}\textbf{c}, \textbf{d}, \textbf{e}).  Informed choices of feature space hold the promise of greatly improving the convergence time, accuracy, and inference of machine-learning algorithms.  Here we show how crucial this choice can be and the roles that our understanding of the underlying biology can play in its selection.  We further demonstrate that this approach identifies key strengths and weaknesses in the theory used to guide the transformations, and thus informs our understanding, or lack thereof, of the underlying biology and physics.

The default choice for feature spaces for our networks would be the centered and standardized raw data\textemdash all vessel radii and lengths for branching networks.  However, theory grounded in evolution, biology, and physics predicts that the parent-to-child ratios of radii and length\textemdash along with associated scaling exponents throughout the networks \cite{west_etal_science_1997, west_etal_nature_1999}\textemdash encapsulate the most biologically-informative properties because they are directly tied to organismic function.  Specifically, numerous models tie these ratios to the ability of branching networks to efficiently fill space and to deliver resources \cite{savage_etal_ploscompbio_2008, west_etal_science_1997, west_etal_nature_1999, savage_etal_pnas_2010, brummer_etal_ploscompbio_2017}.  The fine-scale relationships between fluid flow, global vascular or branching architecture, and vessel or branch morphology are indeed complex  \cite{fluid_mechanics_landau_lifshitz_1987}.  Despite this, much information can be gleaned from the connections between the radial scale factors and hydrodynamics and the length scale factors and space-filling as first-order effects \cite{price_etal_ecolett_2012, bentley_etal_ecolett_2013, newberry_etal_ploscompbio_2015, tekin_etal_ploscompbio_2016}.  As candidates for second-order effects, we also examine branch slenderness exponents.  These couple the radial and length scale factors and inform the likelihood that a branch will experience gravitational buckling under its own weight \cite{smith_etal_newphyt_2014, eloy_etal_natcomm_2017, lopez_etal_jthbio_2011}.

We use recent theory developed by some of us (Brummer, et al.~in  \cite{brummer_etal_ploscompbio_2017}) for the asymmetric branching patterns that are pervasive throughout our data. In this theory the two sibling vessels (labelled $c1$ and $c2$, Figure \ref{fig:tree_pca}\textbf{b}) and the parent vessel (labelled $p$) are combined to give two radial scale factors $\beta_1 = r_{c1}/r_p$ and $\beta_2 = r_{c2}/r_p$ and two length scale factors $\gamma_1 = l_{c1}/l_p$ and $\gamma_2 = l_{c2}/l_p$. Thus the \textit{average} radial and length scale factors are
\begin{equation}
\bar{\beta} = \frac{\beta_1 + \beta_2}{2} \qquad \bar{\gamma} = \frac{\gamma_1 + \gamma_2}{2}
\label{eq:ave_scale_factors}
\end{equation}
To capture sibling branch asymmetry the \textit{difference} radial and length scale factors are
\begin{equation}
\Delta\beta = \frac{\beta_1 - \beta_2}{2} \qquad \Delta\gamma = \frac{\gamma_1 - \gamma_2}{2}
\label{eq:diff_scale_factors}
\end{equation}

Corresponding constraint equations for area-preserving and space-filling branching\textemdash used in canonical optimization models\textemdash are

\begin{align}
(\bar{\beta} + \Delta\beta)^2 + (\bar{\beta} - \Delta\beta)^2 = 1 \label{eq:radial_cons_eq}
\\
(\bar{\gamma} + \Delta\gamma)^3 + (\bar{\gamma} - \Delta\gamma)^3 = 1
\label{eq:length_cons_eq}
\end{align}

Separately, susceptibility to gravitational buckling is quantified in the slenderness exponents \cite{lopez_etal_jthbio_2011}, which relate the scaling of radii to the scaling of lengths as

\begin{align}
\sigma_1 = \frac{\ln(r_{c1}/r_p)}{\ln(l_{c1}/l_p)} \qquad \sigma_2 = \frac{\ln(r_{c2}/r_p)}{\ln(l_{c2}/l_p)}
\label{eq:slenderness}
\end{align}

It is not a priori obvious which combinations of the scale factors will work best as a feature space for discriminating vascular networks. If dynamics of blood flow dominate the formation and evolution of vascular architecture, then variation in scale factors involving vessel radius would be expected to be most informative because vascular theory and empirical evidence show blood flow is most strongly determined by vessel radius.  For example, it is well documented that as blood flow transitions from pulsatile to non-pulsatile, so too does the scaling of vessel radii from the squared scaling (scaling exponent $= 2$) of Eq.~(\ref{eq:radial_cons_eq}) to cubic scaling (scaling exponent $=3$) similar to Eq.~(\ref{eq:length_cons_eq}) \cite{west_etal_science_1997}.  This quantitative shift would then show up as a difference between classified groups in our data that should be detectable by, and informative to, our machine learning algorithms.  If the space-filling constraints and body plan of the organism primarily determine vascular architecture, then variation in scale factors for vessel lengths should best discriminate.  Moreover, average properties might be shared across species while differences or variation around these average properties could reflect distinct selective pressures that can be used to discriminate types of networks and branching principles.  Alternatively, some selective pressures could change the average properties yet share the same values of variation and asymmetry. Finally, if resilience to gravitational buckling determines branching form then the slenderness exponents should differentiate between those organisms susceptible to buckling (plants) and those that are not (mammals).

To test and quantify all of these possibilities, we generate distributions of our data for combinations of the raw and standardized radius and length measurements $(r, l)$ and $(r,l)^\dag$ (where $\dag$ represents the centered and standardized radii and lengths), the slenderness exponents ($\sigma_1, \sigma_2$), and of the symmetric and asymmetric scale factors ($\beta_1$, $\beta_2$, $\gamma_1$, $\gamma_2$, $\bar{\beta}$, $\bar{\gamma}$, $\Delta\beta$, $\Delta\gamma$) for the combined mammal and plant networks.  We first examine the performance of several standard machine-learning techniques to categorize our network data \cite{moses_statistical_modeling_2017, friedman_elements_statistical_learning_2016}.  We use principle components analysis (PCA) to examine feature space variance (Figure \ref{fig:tree_pca}\textbf{f-g}), and compare the results of support vector machine (SVM), logistic regression (LR), and kernel density estimation (KDE) machine-learning methods (Table \ref{tab:features} and Figure \ref{fig:machine}\textbf{a-c}).  Uncertainty is controlled for by graphing the rates of true positive detection versus false positive detection in a one-versus-all comparison between the different classifiers being used while varying the significance of classification (Figure \ref{fig:machine}).  See Supplementary Materials for additional detail on training and testing protocol.  Upon identifying which method has the greatest overall classification success, we then examine which regions in the plant and mammal feature space drive classification and correspond to different species or tissues (Figures \ref{fig:mammal_plant_radius}-\ref{fig:mammal_plant_length}).  Here we account for uncertainty by bootstrapping-with-replacement on the training and testing groups when examining better method at a fixed level of classification significance.  Finally, by drawing on metabolic scaling theory\textemdash the prediction that the scaling of organism metabolism with mass is determined by vascular geometry\textemdash we examine how these different feature spaces constrain variation in estimates of the scaling exponent for organismal metabolic rate (Figure \ref{fig:mse_figure}).

\section*{Results}

\begin{table}
\caption[Global scores for machine learning methods and
 feature spaces.]
{\label{tab:features}\textbf{$|$ Global scores and effect sizes for different machine learning methods and feature spaces in classifying mammal and plant datasets.}}
\resizebox{\textwidth}{!}{
\begin{tabular}{llllllllll}
\hline
& $(r,l)$ & $(r,l)^\dag$ & $(\sigma_1,\sigma_2)$ & ($\beta, \gamma$) & $(\bar{\beta}, \bar{\gamma})$ & $(\Delta \beta, \Delta \gamma)$ & $(\bar{\beta}, \Delta \beta)$ & $(\bar{\gamma}, \Delta \gamma)$ & $(\bar{\beta}, \bar{\gamma}, \Delta \beta, \Delta \gamma)$  \T \B \\
  \cline{2-10}
LogReg  & 0.82 & 0.54 & 0.52 & 0.52 & 0.59 & 0.55 & 0.59 & 0.53 & 0.58  \T \\
SVM  & 0.88 & 0.56 & 0.52 & 0.57 & 0.62 & 0.59 & 0.64 & 0.58 & 0.67 \\
KDE & 2 $\times10^5$**** & 0.049** & 0 & 0.065 & 0.11 & 0.31 & 0.72*** & 0.13* & 0.68****  \B \\
  \hline
\end{tabular}
}
\caption*{\\ \fontsize{10}{8}\selectfont The Logistic Regression and Support Vector Machine scores represent the ratio of correctly classified vessels/nodes for a given feature space, and are compared to a baseline of 0.52 (as determined by the starting ratio of mammal to plant data).  The Kernel Density Estimation scores are test statistic values representing effect size in differentiating mammal from plant networks.  $\dag$ indicates the standardized radius and length distribution.  Asterisks indicate p-value of KDE (no asterisk $p > 0.01$; * $p \le 0.01$; ** $p \le 0.001$; *** $p \le 0.0001$; **** $p = 0$).  All three methods demonstrate high scores (LogReg, SVM) or significant effect sizes (KDE) for the raw radial and length data due to trivial size separation (Figure \ref{fig:tree_pca}\textbf{c}) which is removed upon standardizing for zero mean and unit variance.  Asymmetric scale factor feature space ($\Delta \beta, \Delta \gamma$) is the relatively best performing metric for KDE and LogReg methods, and second best for SVM.  See Figure \ref{fig:machine} for absolute comparison between three methods.}
\end{table}

We demonstrate the importance of choosing theoretically-informed feature spaces over raw data to classify vascular organisms relating form to function.  Classification using only raw data (branch radii and lengths) results only in size-based categorization, an approach that can distinguish between a mouse lung and a Balsa tree, but is not easily applicable to similarly sized organisms or tissues (Figure \ref{fig:tree_pca}\textbf{c}).  Once networks are normalized for size, distributions of the raw data are greatly overlapped \cite{bentley_etal_ecolett_2013, newberry_etal_ploscompbio_2015} (Figure \ref{fig:tree_pca}\textbf{d}) and machine-learning methods applied to the raw data cannot distinguish the networks (Table \ref{tab:features}).  We thus conclude that our theoretically-informed feature spaces are objectively superior at categorizing branching networks over raw data.  In addition, these theoretically-informed feature spaces facilitate much easier translation into known biological principles and constraints on biologic function related to blood flow, space-filling, and metabolic rate.

Importantly, not all theory-motivated features improve classification.  Table \ref{tab:features} shows that the slenderness scaling exponents $(\sigma_1,\sigma_2)$ do no better than random chance.  Two likely explanations are that the definitions in Eq.~(\ref{eq:slenderness}) simultaneously de-couple branching asymmetry and couple radial and length scaling.  This is supported by Figure \ref{fig:tree_pca}\textbf{f-g}, where PCA loadings between $\Delta \beta, \sigma_1$, and $\sigma_2$ are either directly correlated (PCA 2), or are linear combinations of each other (PCA 3 and 4).  Table \ref{tab:features} shows that radial scaling with asymmetry ($\bar{\beta}, \Delta \beta$) outperforms length scaling with asymmetry ($\bar{\gamma}, \Delta \gamma$).  Thus, it appears that transformations that suppress asymmetry and enhance length\textemdash such as the slenderness scaling exponents\textemdash act to obscure the defining features between the mammal and plant data considered.

\begin{figure}
\begin{centering}
\includegraphics[width = 0.8\textwidth]{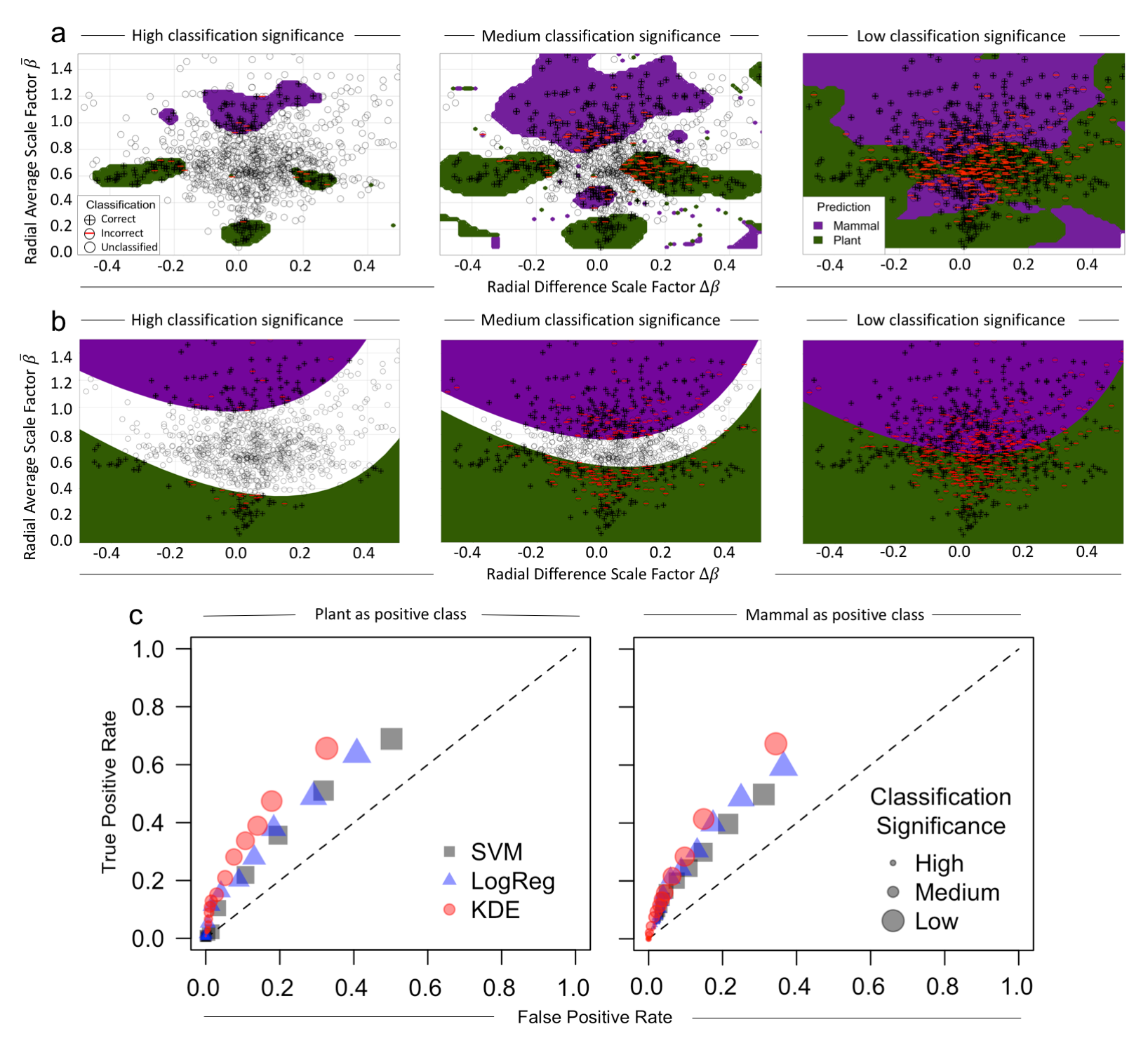}
\caption[Tree diagrams and machine learning results.]{\label{fig:machine} \fontsize{10}{8}\selectfont Comparison of machine learning methods. Results for the \textbf{a} Kernel Density Estimation (KDE) and \textbf{b} Logistic Regression (LR) methods of classification of mammalian and plant networks.  Here, both methods use the radius average and difference scale factors ($\bar{\beta} = (r_{c1} + r_{c2})/2r_p$, $\Delta\beta = (r_{c1} - r_{c2})/2r_p$). For each method, data are randomly split into training (75\%) and testing (25\%) groups.  Following testing, classified points are binned based on predicted probability significance (or score), and comparison is made while varying the level of classification significance from high (left graphs) to low (right graphs).  This procedure was reproduced 100 times, with training and testing division performed at random.  \textbf{c} Receiver operator characteristic (ROC) curves comparing true positive rates (TPR) versus false positive rates (FPR) of classification for methods of Support Vector Machine (SVM), LogReg, and KDE for each level of classification significance.  TPR and FPR are calculated in a one-versus-all framework, where TPR = True Positives$/$(True Positives + False Negatives) and FPR = False Positives$/$(False Positives + True Negatives).  At any given level of significance, three classes exist: mammal, plant, and unclassified.  Thus, the one-versus-all approach means TPR and FPR are calculated separately for either ``mammals and not mammals'' (left graph) or ``plants and not plants'' (right graph).  In both graphs the KDE method is shown to outperform the LogReg and SVM.}
\end{centering}
\end{figure}

\begin{figure}
\begin{centering}
\includegraphics[width = 0.75\textwidth]{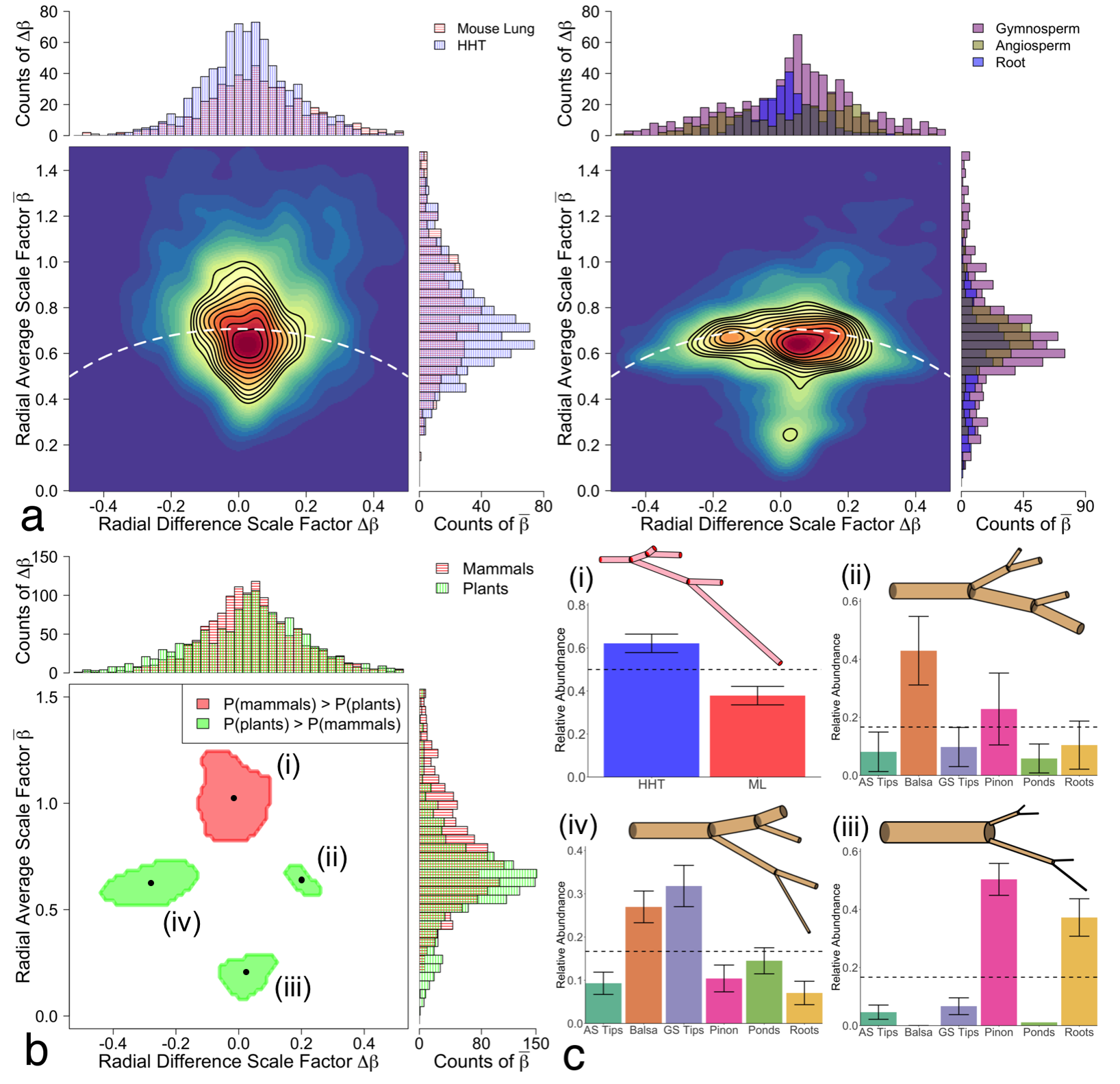}
\caption[Bivariate distribution graphs of mammal and plant radial and length scale factors.]{\label{fig:mammal_plant_radius} \fontsize{10}{8}\selectfont Classification based on features  (the radial scale factors $\bar{\beta}$ and $\Delta \beta$) that are related to fluid transport--blood or sap--via volume-flow rate and hydraulic resistance through networks and vessels.  \textbf{a}, Joint and marginalized distributions for the mammal (left) and plant (right) radius scale factors using the KDE method.  Mammals are divided into mouse lung and HHT, and plants are divided into the groups of Gymnosperms (GS), Angiosperms (AS), and Roots.  Black contours represent lines of constant probability density, ranging from 0.5 to 0.05 in steps of 0.05.  White dashed lines are graphs of the radius conservation equation for area-preservation.  \textbf{b}, Regions of significantly ($p < 0.05$) greater joint probability density for the mammals (red) or plants (green).  \textbf{c}, Representative diagrams of tree networks and bar plots of relative abundances of each group/species are presented for each region of significant classification in \textbf{b} (clockwise).  Scale factor values for tree networks are determined by geometrically averaging over all classified data points within each significance region.  Means and standard deviations for bar plots are determined by bootstrapping the KDE method 1000 times.  Horizontal black dashed lines represent null expectations of relative abundances. These significance-region abundances are corroborated with global-level testing of all pairs of branching networks (See Supplementary Table S1).  The global-level test is a method that effectively integrates over the entire feature space to produce one singular $p$-value for the comparison \cite{duong_etal_pnas_2012}.}
\end{centering}
\end{figure}

Differences in machine learning performance across methods (KDE, SVM, LR) is due to the multivariate structure of the feature spaces being studied and the chosen machine learning method.  In particular, the KDE method excels at resolving the multimodality \cite{duong_jnonparstats_2013} that characterizes the radial scale factors for the plant dataset (Figures \ref{fig:machine}\textbf{a} and \ref{fig:mammal_plant_radius}\textbf{a-b}).  Since the distribution means are approximately equivalent, the SVM and LR methods are strongly influenced by outliers and the higher moments comprising the mammal dataset (Figures \ref{fig:machine}\textbf{b})  \cite{friedman_elements_statistical_learning_2016}. 

Comparing across all methods and features spaces, we find that the combination of the KDE method and the average and difference scale factors for radius ($\bar{\beta}, \Delta\beta$) are the most effective for classifying branching network data (Figure \ref{fig:machine}, Table 1, and Supplementary Material Tables S1-S3). The fact that variation in the feature space for radial scale factors is the best has strong implications about what functional features form the major distinctions between biologic networks.  \textit{Specifically, our empirical finding of the primacy of information based on scaling ratios of radii strongly suggests that hydrodynamic principles are the primary drivers of vascular branching patterns and overall network form.}  

Multiple theories of vascular networks, as well as basic physics and fluid mechanics, dictate that rates of fluid flow are largely governed by the total cross-sectional area of vessels or limbs, which can be exactly related to the ratios of scaling radii used in our feature space \cite{savage_etal_ploscompbio_2008}.  Importantly, theory recently developed by us demonstrates that there can exist a range of morphologies that still adhere to these area-preserving\textemdash pulsatile flow in mammals or external branching in plants\textemdash or area-increasing predictions\textemdash non-pulsatile flow in mammals \cite{brummer_etal_ploscompbio_2017}.  In contrast, variation in the ratios of vessel lengths is more strongly tied to the ability of the vascular network to fill the body.  Thus, length ratios appear to either encapsulate little important information about the differences among biologic networks, or they may not adequately capture the key properties of space filling for the architecture of vascular networks \cite{bentley_etal_ecolett_2013, newberry_etal_ploscompbio_2015, brummer_etal_ploscompbio_2017, hunt_etal_pre_2016}.  Having identified the best performing feature space and machine learning method, we now delve deeper into the variation in the architecture and functional properties of vascular networks. 

Focusing on the KDE method we see that mammalian branching exhibits more area-increasing branching than plants (Figure \ref{fig:mammal_plant_radius}\textbf{b(i)}).  Area-increasing branching is necessary to simultaneously increase total surface area for oxygen and metabolite transport and to slow blood flow as it travels from the heart to the capillaries and transitions from pulsatile to non-pulsatile flow, the latter phenomenon not present in plants.  However, values of $\bar{\beta} \approx 1.0$ and $\Delta\beta \approx 0$ represent a deviation from the theoretical predictions of $\Delta\beta = 0$ and $\bar{\beta} = 1/2^{1/3} \approx 0.794$ for the non-pulsatile flow expected in this region.  This marked increase in cross-sectional area is shared by both the HHT and ML networks as indicated by the nearly null relative abundances of these two networks (Figure \ref{fig:mammal_plant_radius}\textbf{c(i)}) as well as by the insignificant $p$-value score of 0.2 from the global-level implementation of the KDE method (See Supplementary Table S1).  This suggests that transitions in blood flow type from pulsatile to non-pulsatile may occur across a greater range of branching generations, and begin nearer to the heart, than in current theory \cite{savage_etal_ploscompbio_2008, kolokotrones_etal_nature_2010}.

The majority of plant networks adhere to area-preservation while exhibiting a greater tendency than mammals to branch asymmetrically (specifically the Balsa, Pi\~{n}on, Ponderosas, and GS Tips, Figure \ref{fig:mammal_plant_radius}\textbf{a} and \textbf{b}).  Within the plants we find that differentiation is driven at the species level (Figure \ref{fig:mammal_plant_radius}\textbf{c(ii)-(iv)} and Supplementary Table S1), unrelated to plant categorization as angiosperm or gymnosperm.  For example the Balsa, an angiosperm, is the only species present in both the positive and negative asymmetry types (Figures \ref{fig:tree_pca}\textbf{b}) as demonstrated by being the only network with its standard deviation outside the null expectation in Figure \ref{fig:mammal_plant_radius}\textbf{c(ii)} and \textbf{(iv)}.  Thus, the Balsa consists of two unique branching motifs that distinguish it from GS Tips, and the Pi\~{n}on and Roots that have large relative abundances in one region each\textemdash the negative asymmetric branching of motif \textbf{c(iv)} and the symmetric branching of motif \textbf{c(iii)}, respectively.

Mechanisms for the asymmetry and motifs observed in the plant radial scale factors are likely due to functional trait plasticity associated with light-seeking behavior, self- and wind-induced pruning, gap-filling, and other environmental stressors.  However, making quantitative connections remains an open challenge \cite{smith_etal_newphyt_2014, eloy_etal_natcomm_2017, lopez_etal_jthbio_2011}.  For example, the slenderness scaling exponents can be calculated for all six scenarios in Figure \ref{fig:mammal_plant_radius}\textbf{c(ii)-(iv)}, beginning with the first generation of child branches.  Expressing Eq.~(\ref{eq:slenderness}) in terms of the average and difference scale factors

\begin{align}
\sigma_1 = \frac{\ln(\bar{\beta} + \Delta \beta)}{\ln(\bar{\gamma} + \Delta \gamma)} \qquad \sigma_2 = \frac{\ln(\bar{\beta} - \Delta \beta)}{\ln(\bar{\gamma} - \Delta \gamma)} 
\end{align}

Thus, the slenderness scaling exponents for each plant motif are: $\sigma_1 = -12.3, \sigma_2 = 1.4$ for motif (ii); $\sigma_1 = -19.0, \sigma_2 = 5.1$ for motif (iii); and $\sigma_1 = -4.8, \sigma_2 = 0.54$ for motif (iv).  Biomechanical theory that applies columnar (Euler beam) buckling to branching systems demonstrates that slenderness exponents of $\sigma \ge 1$ are structurally advantageous for plant architectures as they push the locations of breakage points into the canopy as opposed to the trunk.  Yet, the slenderness exponents we calculate for the observed motifs do not entirely agree with this framework, despite adherence of the radial scale factors in motifs (ii) and (iv) to the area-preserving branching constraint of Eq.~(\ref{eq:radial_cons_eq}).  This disparity may lie in two sources: (i) the slenderness exponent formula of Eq.~(\ref{eq:slenderness}) was originally derived using symmetrically branched networks and (ii) the length scaling exponents involved in Eq.~(\ref{eq:slenderness}).  The latter issue we now investigate.

\begin{figure}
\begin{centering}
\includegraphics[width = 0.75\textwidth]{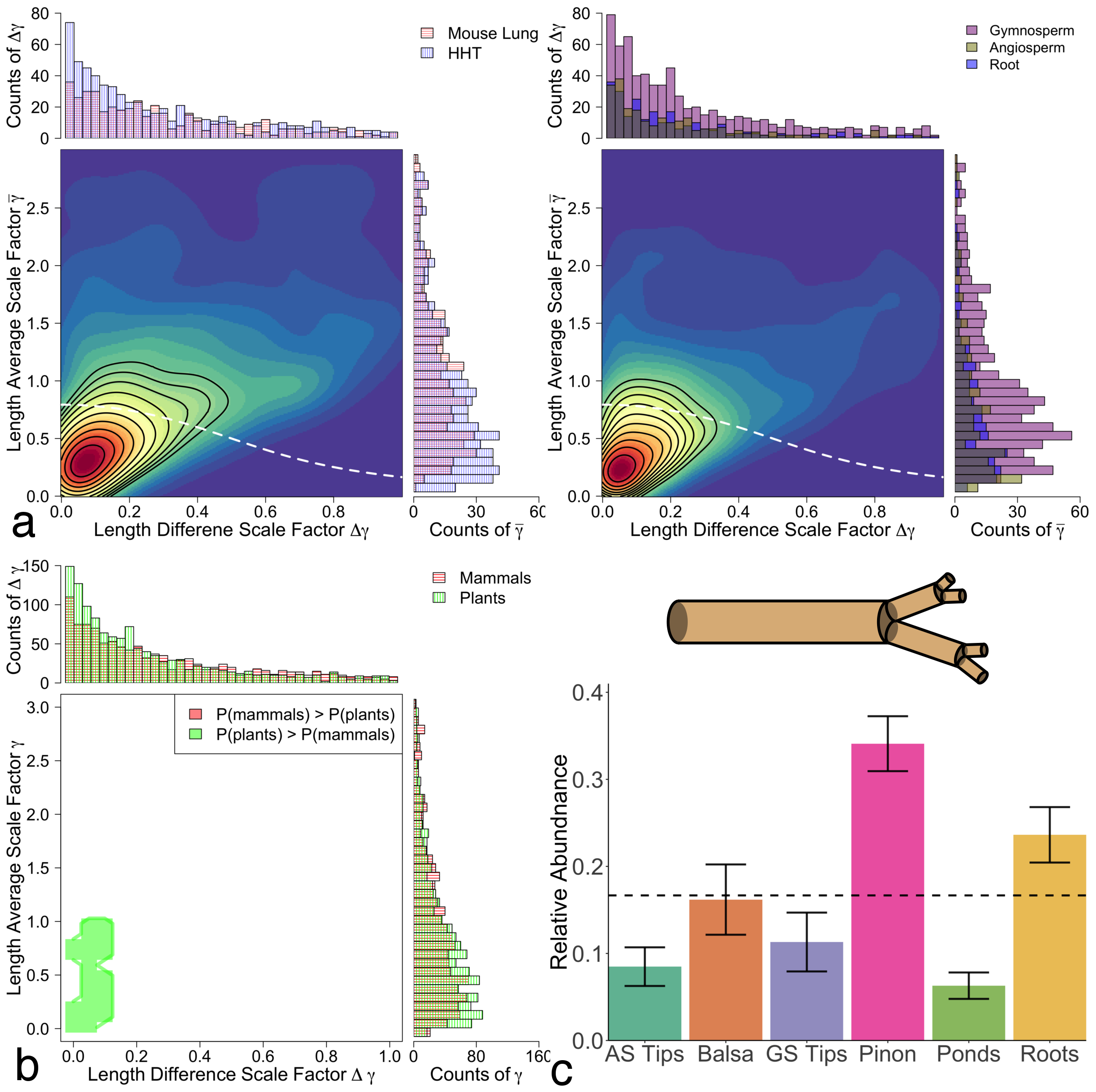} 
\caption{\label{fig:mammal_plant_length} \fontsize{10}{8}\selectfont  Classification based on features (the length scale factors $\bar{\gamma}$ and $\Delta \gamma$) that are related to costs of materials and construction for these networks as well as the extent to which they fill the space of the organisms that they are supplying with nutrients and resources. \textbf{a-c}, See caption for Figure \ref{fig:mammal_plant_radius} for description of subfigures.  The significance-region abundances in \textbf{c} are corroborated with global-level testing of all pairs of branching networks (See Supplementary Table S3)}
\end{centering}
\end{figure}

Connecting length-based categorization to mechanism\textemdash the space-filling constraint of Eq.~(\ref{eq:length_cons_eq}) \textemdash remains a challenge.  The combination of the KDE method and length scale factor feature space ($\bar{\gamma}, \Delta\gamma$) identified only one region of significance. In this region differentiation is driven by the plants, specifically the Pi\~{n}on and Roots (Figure \ref{fig:mammal_plant_length} and Supplementary Table S3).  This is despite the large amount of variance explained by the length scale factors in the principle components analysis (Figure \ref{fig:tree_pca}).  The single region driving differentiation corresponds to average length scale factors $\bar{\gamma} < 1$.  This effect would normally result in an increase in the slenderness exponent, Eq.~(\ref{eq:slenderness}), driving gravitationally induced buckling (self-pruning) to occur in the canopy instead of at the trunk ($\sigma \ge 1$) \cite{lopez_etal_jthbio_2011}.  However, median values of $\sigma$ for both the plants and mammals were approximately 0.2, far below the needed theoretical threshold of $\sigma = 1$.  We interpret this deviation from expected biomechanics as an indicator that the length scale factors, as defined, are poor features for characterizing vascular or branching architecture.

The inability of the length scale factors to inform classification between networks suggests several scenarios.  Two contrasting and extreme scenarios are that either a universal architecture or a completely random architecture is being followed by both the mammals and plants \cite{tekin_etal_ploscompbio_2016}.  This result is unlike the radial scaling that is strongly coupled to hydraulics.  Current theory suggests that the architecture associated with length scaling is guided by the principles of space-filling fractals \cite{savage_etal_ploscompbio_2008, west_etal_science_1997, west_etal_nature_1999, brummer_etal_ploscompbio_2017}.  However, large deviations are observed between the joint distributions of the length scale factors and the theoretical curves determined by the space-filling conservation equation (Figure \ref{fig:mammal_plant_length}\textbf{a}). A third scenario is that there exists a disconnect between how length scale factors are conventionally defined in simplified models versus how they are measured in complicated natural systems.  All three scenarios support the need for including missing constraints, variables, and assumptions (e.g. branching angles, multi-fractal scaling, etc.), or alternative mathematical frameworks \cite{hunt_etal_pre_2016, warner_bullmathbio_1976, banavar_etal_pnas_2010, dodds_prl_2010, barnsley_fractals_2014}.

To better understand the physiological and biological implications of these categorizations, we examine the influence of asymmetric branching on estimates of biological rates\textemdash specifically the metabolic scaling exponent $\theta$ that canonically relates metabolic rate $B$ to body mass $M$ as $B \propto M^\theta$.  Previous studies spanning orders of magnitude in body mass have shown that $\theta$ converges on a value near 3/4, yet exhibits variation specific to mammals or plants \cite{savage_etal_funceco_2004, mori_etal_pnas_2010, kolokotrones_etal_nature_2010}.

\begin{figure}
\begin{centering}
\includegraphics[width = 0.75\textwidth]{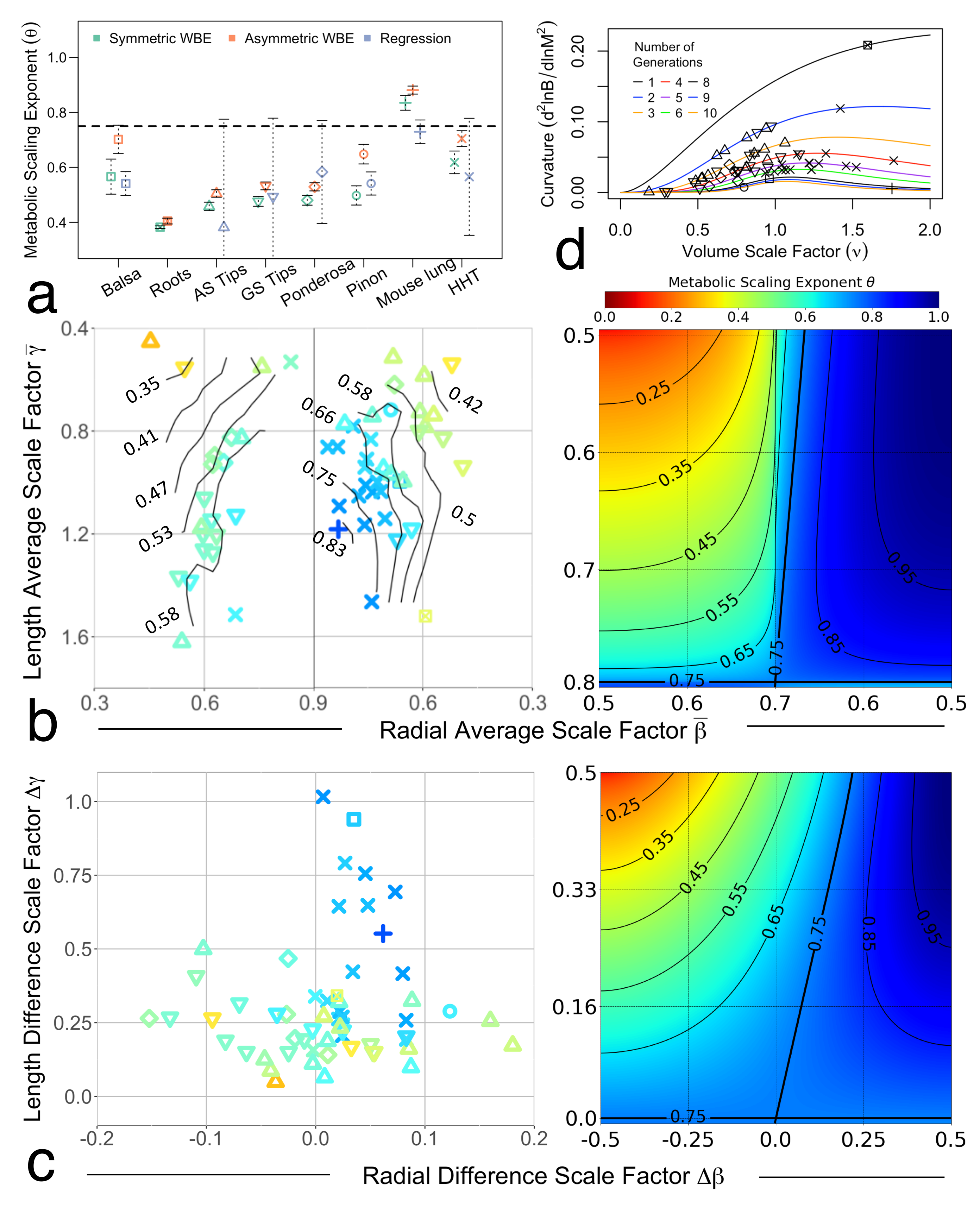}
\caption[Heatmaps of empirically measured metabolic scaling exponent values.]{\label{fig:mse_figure} \fontsize{10}{8}\selectfont Variation in metabolic scaling exponents related to variation in branching geometry. \textbf{a} Comparison of \textit{symmetric} (red) and \textit{asymmetric} (green) estimates of metabolic scaling exponents to regression (blue) based estimates.  For groups with multiple species and/or multiple individuals (AS Tips, GS Tips, Ponderosa, and HHT), metabolic scaling exponents were calculated at the species/individual level when averaged.  Error bars represent $95\%$ confidence intervals.  The horizontal dashed line represents a metabolic scaling exponent value of 3/4. Note that \textbf{a} also serves as a legend for the symbols in all other subfigures.  \textbf{b}, Empirically based estimates of metabolic scaling exponents are presented as functions of the geometrically averaged length and radial average scale factor values (left), and compared to theoretical predictions (right) reproduced from Brummer et al.~(ref.~10).  \textbf{c}, Analogous results are presented as in \textbf{b} but instead for the length and radius difference scale factors.  Solid black lines represent contours of constant metabolic scaling exponent values.  Axis ranges differ between empirical data and theory-based predictions due to observed deviations from conservation equations.  \textbf{d}, Curvature of metabolic rate versus mass (log-log) as a function of volume scaling ($\nu_{\ASYM}$) and the number of branching generations (N).}
\end{centering}
\end{figure}

To probe this variation we use branching data to estimate metabolic scaling (Figure \ref{fig:mse_figure}) by directly accounting for network geometry and size \cite{savage_etal_ploscompbio_2008, west_etal_science_1997, brummer_etal_ploscompbio_2017},
\begin{equation}
\theta = 
	\frac{\ln(2^N)}{\ln(2^N) + \ln(1 - \nu^{N+1}) - \ln(\nu^N(1-\nu))} 
\label{eq:mse_exact}
\end{equation}
where $N$ is the total number of branching generations in the network and $\nu$ represents volumetric scaling\textemdash the ratio of the sum of the volumes of both child branches to the volume of the parent branch.  Specification of $\nu$ allows estimation of $\theta$ under different model assumptions for symmetric ($\nu = 2\beta^2\gamma$) or asymmetric ($\nu = 2 \bar{\beta}^2\bar{\gamma} + 4 \bar{\beta}\Delta\beta\Delta\gamma + 2\bar{\gamma}\Delta\beta^2$) branching.  We also use a regression method between the number of terminal branches $N_{\TIPS}$ and total volume $V_{\TOT}$ distal to a given branch ($N_{\TIPS} \propto V_{\TOT}^\theta$) that does not depend directly on geometry (See Supplementary Materials).
%
%
We find that asymmetric branching increases the predicted values of metabolic scaling exponents when compared to the symmetric- and regression-based methods (Figure \ref{fig:mse_figure}\textbf{a}).  This is due to all networks exhibiting some length asymmetry, and more importantly suggests that previous studies have underestimated metabolic scaling exponents by not accounting for such variation \cite{bentley_etal_ecolett_2013, huo_etal_jrsi_2012, lau_etal_forestecomanage_2019}.

To understand which different scale factors are primarily responsible for observed variation in the predicted metabolic scaling exponents we focus on the asymmetric version of Eq.~(\ref{eq:mse_exact}).  Estimated metabolic scaling exponents are graphed for each individual organism in terms of the average scale factors ($\bar{\beta}, \bar{\gamma}$) in Figure \ref{fig:mse_figure}\textbf{b} and difference scale factors ($\Delta \beta, \Delta \gamma$) in Figure \ref{fig:mse_figure}\textbf{c}.  We compare these graphs against the corresponding theoretical predictions reproduced from Brummer et al.~in  \cite{brummer_etal_ploscompbio_2017} where we have graphed the approximate form of Eq.~(\ref{eq:mse_exact}),
\begin{equation}
\theta \approx \frac{\ln(2)}{\ln(2) - \ln(2\bar{\beta}^2\bar{\gamma}) - \ln(1 + \frac{2\Delta\beta\Delta\gamma}{\bar{\beta}\bar{\gamma}} + \frac{\Delta\beta^2}{\bar{\beta}^2})}
\label{eq:mse_asym}
\end{equation}
assuming small volume scaling ($\nu < 1$), generationally large networks ($N >> 1$), and enforcing area-preserving and space-filling (Eqs.~(\ref{eq:radial_cons_eq}) and (\ref{eq:length_cons_eq})).

We observe a striking amount of grouping among the mammals and plants when graphing the metabolic scaling exponent $\theta$ versus the average radial and length scale factors $\bar{\beta}$ and $\bar{\gamma}$ (Figure \ref{fig:mse_figure}\textbf{b}).  This indicates that, of all the features and data considered, the average scale factors ($\bar{\beta}$, and $\bar{\gamma}$) are the primary determinants of variation in the metabolic scaling exponent and thus organism function.  

In contrast to previous theory and importantly for understanding how diverse branching architectures could lead to universal scaling exponents, we find near constancy of the metabolic scaling exponent despite large fluctuations in length scaling (Figure \ref{fig:mse_figure}\textbf{c}).  These shared exponents are likely driven by the little to no radial asymmetry observed in mammalian networks and suggests that variation in length asymmetry ($\Delta \gamma$) in vascular networks has little influence on whole organism metabolic function in the presence of symmetric radial branching ($\Delta \beta = 0$).

Figures \ref{fig:mse_figure}\textbf{b-c} demonstrate marked deviation in the observed grouping (or lack thereof) between the empirically based predictions of metabolic scaling from Eq.~(\ref{eq:mse_exact}) and the constraint-based theoretical predictions of metabolic scaling from Eq.~(\ref{eq:mse_asym}).  To explore this deviation we calculate curvature between metabolic rate and mass in log-log space (Supplementary Materials).  When branching networks are strictly assumed to be very large, ($N >> 1$) and decreasing in volume in all segments across any generation ($\nu < 1$, Eq.~(\ref{eq:mse_asym})), we predict zero curvature, regardless of the extent of branching asymmetry.  When accounting for variation in network size and volume scaling (Eq.~(\ref{eq:mse_exact})) we predict positive (concave up) curvature (Figure \ref{fig:mse_figure}\textbf{d}).  These predictions are both in agreement with respiration-based studies of mammals \cite{kolokotrones_etal_nature_2010}, and demonstrate the need for theories of metabolic scaling that incorporate the finite-size of the network.  Furthermore, we predict that curvature decreases to zero with increasing network size, or generation $N$, in agreement with respiration-based studies of plants \cite{mori_etal_pnas_2010}.   These results can be informative for future studies that simultaneously connect branching patterns and vascular data to ontogenetic- and size-based shifts in organismal metabolism.  Such shifts are observed in growth and reproduction curves for tumors \cite{herman_etal_plosone_2011}, plants \cite{mori_etal_pnas_2010}, mammals \cite{west_etal_nature_2001}, and fish \cite{barneche_etal_science_2018}.

\section*{Discussion}

Machine learning is a powerful tool, but often considered a black box.  We show that by combining machine learning with mechanistic theory it can be made more effective and provide insight into physiological mechanism.  Here we take a first step towards building that bridge by using mechanistic theory of vascular networks to choose better feature spaces.  In so doing, we achieve a two-fold, mutually reinforcing benefit: 1.~We achieve better results for categorizing networks than if we used either the raw feature space or a mechanistically inspired feature space that predicts only one morphology (the symmetric scale factors $\beta$ and $\gamma$), and 2.~Results are much more interpretable.  For example, the best-performing feature spaces\textemdash the asymmetric ratios of vessel and limb radii, $\bar{\beta}$ and $\Delta\beta$\textemdash are explicitly and naturally tied to specific mechanisms\textemdash hydrodynamic constraints and resource flow\textemdash and allow for variation in form while still following these constraints.  Additionally, the under-performing feature spaces\textemdash ratios of vessel and limb lengths, $\bar{\gamma}$ and $\Delta\gamma$\textemdash identify what may be potential weaknesses in the theory and avenues for new development to provide greater specificity.  Alternatively, the inability of the ratios of vessel and limb lengths to classify between mammals and plants may be pointing to broadly shared architectural principles that are not specific to mammals or plants.  This is despite different mechanistic demands, such as structural support for plants.

The results of this study also serve to inform our understanding of the physiological pressures that determine convergence in organismal form and function. We find that variation in vascular based estimates of metabolic scaling exponents\textemdash in particular curvature\textemdash is primarily determined by variation in the average scale factors ($\bar{\beta}$ and $\bar{\gamma}$), symmetric radial branching, and relative network size.  This result helps to resolve some of the contradictory size-based observations in variation in metabolic scaling between mammals and plants \cite{mori_etal_pnas_2010, kolokotrones_etal_nature_2010}.  It emphasizes the needs to develop models of vascular networks that can better account for finite-size measurements (clustered sampling versus whole network measurement), to acquire comprehensive datasets that span the entirety of the vascular branching structures being studied, and to simultaneously acquire respiration based measurements of organismal metabolism.

In this direction, a shortcoming of our model of metabolic curvature is its complete inability to capture negative curvature.  This scenario arises when examining individual growth curves in mammals, plants, and tumors when the metabolic scaling exponent decreases from linear to sub-linear (typically from 1 to 3/4) \cite{mori_etal_pnas_2010, herman_etal_plosone_2011, west_etal_nature_2001}.  Failure to capture this essential biologic feature should spur continued development of theories of metabolic scaling and vascular branching.

Finally, this work has implications for several fields, spanning bio-mechanical and physiological imaging and theory to machine learning and biomedical applications.  Incorporating topological features\textemdash connectivity and loops\textemdash and branching angles could enhance categorization methods because these features provide structural integrity and redundancy to damage in plant leaves and in capillaries \cite{mileyko_etal_plosone_2012, katifori_etal_prl_2010, zamir_jgenphys_1978}.  Additional measures that capture organ and organismal physiology could provide further insight and tests.  Examples for mammalian tissues include flow reserve\textemdash the change in blood flow between normal and dilated vessel states\textemdash or blood perfused for a given vascular tree \cite{ohuchi_etal_pediares_2007, choi_etal_physioreports_2020}.  New applications of tomographic imaging and computer vision techniques to plants\textemdash light detection and ranging (LiDAR) and positron emission tomograpahy (PET)\textemdash are greatly expanding digitized plant architecture datasets and allowing for the direct inclusion of branch angles and xylem and phloem transport measurements as part of the biological feature space \cite{lau_etal_trees_2018, lau_etal_forestecomanage_2019, wang_etal_physmedbio_2014, hubeau_etal_trendsplantsci_2015}.  Simultaneously, advances in medical imaging and vascular segmentation algorithms are leading to datasets of fully connected branching and blood vessel networks \cite{moccia_etal_compmethprogbiomed_2018}.  Such expansive datasets previously unavailable will allow for comprehensive testing of vascular branching theories where, in principle, machine learning based motif identification could be used to digitally regenerate branching networks using iterated function systems \cite{warner_bullmathbio_1976, barnsley_fractals_2014}.

Utilizing more robust applications of machine-learning methods and model complexity might help improve classification based on raw data and should improve classification using feature spaces based on theory as well.  In closing, this work provides a proof-of-principle that a mechanistically-based automatic classification and detection scheme for vascular networks could have application in medical diagnostics for long term progressive disease (e.g., tumor growth).  Here, classification is driven by outlier detection between vascular networks surrounding and comprising tumors compared against verified healthy vascular neteworks \cite{wang_etal_lungcancer_2017}. Such an application would serve as a new dimension in radiomic studies where the detection and classification of tumors based on vascular branching is wholly absent \cite{alilou_etal_scireports_2018, lambin_etal_natrevcliniconc_2017} and could provide an alternative measure of tumor growth and development \cite{pashayan_etal_science_2020}.




%
\bibliographystyle{vancouver}

\section*{Funding}This work was supported by the National Science Foundation Grant 1254159
\section*{Author Contributions} ABB, BJE, and VMS conceived the analysis, ET, LPB, VB, AG, and IO acquired the data.  ABB, PL, and JS, analyzed the data.  ABB and VMS developed theory.  ABB, BJE, and VMS wrote the manuscript.  All authors edited the manuscript.
\section*{Competing Interests} The authors declare that they have no
competing financial interests.
\section*{Correspondence} Correspondence and requests for materials
should be addressed to ABB~(email: abrummer@ucla.edu).
\section*{Data Availability} The datasets and code that support the findings of this study have been uploaded as part of the supplementary material.



\section*{Supplementary materials}
Supplementary Text\\
Figures S1 to S3\\
Tables S1 to S6\\
References 1-19\\
Supplementary Datasets\\
Supplementary Code

%
%
%
%
%
%
%
%
%
%

%
%
%

\captionsetup[table]{labelfont={bf}, labelformat={default}, labelsep=space, name={Table S\hspace{-1mm}}}
\captionsetup[figure]{labelformat={default}, labelsep=colon, name={Figure S\hspace{-1mm}}}
\captionsetup[table]{labelfont={bf}, labelformat={default}, labelsep=space, name={Table S\hspace{-1mm}}}
\renewcommand{\theequation}{S.\arabic{equation}}
\renewcommand*{\citenumfont}[1]{S#1}
\renewcommand*{\bibnumfmt}[1]{[S#1]}
%
%
%
%
%
\appendix

\newpage
{\bf SUPPLEMENTARY INFORMATION}
\setcounter{table}{0}
\setcounter{figure}{0}
\setcounter{equation}{0}
\setcounter{section}{0}

\tableofcontents

\newpage

\section{Classifier methods}
To generate probability models using the support vector machine and logistic regression methods, approximately 75\% of each combined plant and and combined mammal dataset was taken as the training sample and the remaining 25\% used as the testing sample.  This was done after randomization within respective datasets to minimize the chances of accidentally removing an entire individual or species.  Then, using the radial scale factor feature space ($\bar{\beta}$ and $\Delta \beta$) the two groups of mammals and plants were compared using three methods: a support vector machine (SVM) model and a logistic regression (LR) model, and a non-parametric kernel density estimator (KDE) method.

The support vector machine method, a supervised machine learning algorithm for classification problems, plots data points in an n-dimensional space and draws a decision boundary by maximizing the margin between points from different classes. To generate a support vector machine model for our datasets, we used the SVC (support vector classification) function from the Python scikit-learn package. When running SVC, we used the polynomial kernel with a degree of 2.

The LR method generates probabilities of how likely certain data points are going to fall under a certain class, based on the logit function $1/(1 + e^{-x})$. To produce a non-linear LR model for our two-dimensional datasets, we used the Python scikit-learn logistic regression function and added three nonlinearity columns, $x_1^2, x_2^2$, and $x_1\times x_2$. All pairwise combinations were run on this LR model and the probabilities were recorded.

For both the SVM method and the LR method, a polynomial kernel was used. Through a series of tests between different kernels, including linear and radial basis function, the polynomial kernel yielded the highest accuracy score when classifying the testing data.  The LR and SVM methods differ in their nonlinearities, where the LR method uses an added nonlinearity term and the SVM method uses a radial basis function kernel. The probabilities for the LR method are assigned using the logit function, and the probabilities assigned using the SVM method are based on the point distances from the decision boundary.  For both of these methods, testing points are classified based on the value of their score.  When using training data that is equally split between the two categories then points that receive scores of 0.5 or greater are classified as one group, and scores below 0.5 are classified as the other group.

The kernel density estimator (KDE) technique introduced by Duong et al.~\cite{s_duong_etal_pnas_2012, s_duong_jnonparstats_2013} generates non-parametric, multi-dimensional probability distributions, $\mathbb{P}_i(\textbf{x})$, of vascular traits, $\textbf{x}$, for each testing group, $i = A, B$.  These distributions are then compared against one another for their extent of uniqueness, or non-overlap, represented by the test statistic $T = \int[\mathbb{P}_A(\textbf{x}) - \mathbb{P}_B(\textbf{x})]^2d\textbf{x}$.  This test can be thought of as a multi-dimensional generalization of the two-sample Kolmogorov-Smirnov (KS) test, where significance of classification is conventionally communicated through \textit{p}-values.  We use the nominal threshold of $p \le 0.05$ as a threshold for significance when using the KDE method.

The KDE method can be applied both globally \cite{s_duong_etal_pnas_2012} and locally \cite{s_duong_jnonparstats_2013}.  At the global level, the test statistic $T$ is calculated as described above (and in more detail in \cite{s_duong_etal_pnas_2012}), and converted to a $p$-value using standard tables.  While $p = 0.05$ is used as the nominal threshold for significance, it should be noted that $p$ values ranging orders of magnitude in size from $p = 10\times10^{-2}$ to $p = 10\times10^{-20}$ are observed, and thus interpreted as relative levels of significance.  The local application of the KDE method is effectively an inversion of the calculation for the test statistic $T$.  Here, one sets the desired threshold for significance, or the value of the test statistic $T$, then calculates which regions in the vascular trait-space, $\textbf{x}$ correspond to the chosen test statistic.

\newpage

\section{Feature space selection}

Here we describe our methods for selecting the feature space.  We tested each machine learning method on a variety of vascular trait feature spaces.  Specifically, we tested the feature spaces of: raw and standardized diameter and length measurements; the slenderness scaling exponents, $\sigma_1$, and $\sigma_2$; the symmetric WBE diameter and length scale factors, $\beta$ and $\gamma$; and five combinations of the asymmetric scale factors.  The five combinations of asymmetric scale factors were: the average scale factors $\bar{\beta}$ and $\bar{\gamma}$; the difference scale factors $\Delta \beta$ and $\Delta \gamma$; the radial scale factors $\bar{\beta}$ and $\Delta \beta$; the length scale factors $\bar{\gamma}$ and $\Delta \gamma$; and all four asymmetric scale factors $\bar{\beta}, \bar{\gamma}, \Delta \beta,$ and $\Delta \gamma$.  These results are presented in Table \ref{tab:features}.  We also conducted a principle components analysis (PCA) on the asymmetric scale factors to identify which combinations of scale factors explain the most variance in the datasets (Figure \ref{fig:tree_pca}\textbf{f-g}).  Performing a PCA on all vascular variables is non-trivial due to the non-random presence of empty-cells in the dataset \cite{s_dray_etal_plantecol_2015}.

\subsection{Raw radius and length measurements}
Prior to transforming the radial and length measurements of our vascular datasets as motivated by scaling theory \cite{s_newberry_etal_ploscompbio_2015, s_bentley_etal_ecolett_2013,s_brummer_etal_ploscompbio_2017}, we applied all three of the logistic regression (LR), support vector machine (SVM), and kernel density estimator (KDE) methods on the raw, untransformed data.  This was done in two ways, first on the data after all measurements were converted to \textit{meters} (Figure \ref{fig:tree_pca}\textbf{c}), then again after performing a standardized transformation by translating each species- or group-level distribution to be centered about zero, and then normalizing by the respective standard deviations (Figure \ref{fig:tree_pca}\textbf{d}).

Classifying vascular data based on metric size is both trivial and uninformative.  Small networks (mouse lung) are clearly distinct from large networks (whole trees), and varying degrees of overlap will exist in the intermediate range of all other networks (Figure \ref{fig:tree_pca}\textbf{c}).  All three methods yield significant global classification scores, as demonstrated by the example scores of LR: 82\%, SVM: 88\%, and KDE: T-stat $\approx 2\times10^5$ found in Table \ref{tab:features}.  One can certainly argue that classifying networks in this manner is possible, even though it is simply demonstrating that networks of differing size are distinguishable.  However, this approach does not provide an obvious path toward understanding at a mechanistic level why certain patterns are observed beyond the simple size-based classification.  Furthermore, when applying these methods for the purposes of disease detection one is oftentimes examining healthy and diseased tissues that are of similar size classes.  Thus, the utility of size-based classification is rendered moot, and we must transform.  



The most common transformation is to standardize the data such that the distribution means are centered about the origin and to normalize by the variance \cite{s_friedman_elements_statistical_learning_2016, s_moses_statistical_modeling_2017} (Figure \ref{fig:tree_pca}{\textbf{d}).  While this approach has the benefit of removing the global size-based hierarchy between the networks, it fails to address the common pattern of local size-based hierarchy that is commonly found within networks \cite{s_newberry_etal_ploscompbio_2015, s_bentley_etal_ecolett_2013, s_tekin_etal_ploscompbio_2016}.  Specifically, the abundance-size distribution of vessels in a vascular network is approximately exponential due to the fact that the vessels, on average, decrease in size at every bifurcation.  Thus, classification becomes immediately obscured, as demonstrated by the noticeably decreased example scores of LR: 54\%, SVM: 56\%, and KDE: T-stat $= 0.049$ (with a corresponding $p$-value less than 0.001) in Table \ref{tab:features}.  The fact that the KDE method still yielded a significant score is a consequence of this method excelling at detecting multimodality between distributions.  Yet, even though the KDE method does yield a significant score, the effect size is negligible and we still have the original problem of how to interpret the results, only now it is further compounded after having performed the standardization transformation.  It is possible to consider alternative transformations based on mechanical principles\textemdash flow-rates and pressures \cite{s_yang_etal_amjheartphys_2010}\textemdash yet the problem of the hierarchy of sizes returns, only now with different physical units.  As a consequence of these complications in analyzing raw data, we turn to classification based on scale factor variables as guided by the metabolic scaling theory literature.

\subsection{Scale factor feature spaces}
To investigate the scale factor feature spaces, all three machine learning methods were tested on different combinations of candidate feature spaces, and a principle component analysis (PCA) was conducted on the asymmetric scale factor feature space.  Results from the application of different machine learning methods are presented in Table \ref{tab:features} and graphs from the PCA are presented in Figure \ref{fig:tree_pca}\textbf{f-g}.  Table \ref{tab:features} shows that the top two combinations of features for classifying animal and plant vascular networks are the set of all asymmetric scale factors ($\bar{\beta}, \Delta \beta, \bar{\gamma}, \Delta \gamma$) and the radial scale factors ($\bar{\beta}, \Delta \beta$) .  That the set of all asymmetric scale factors performs the best is not surprising as this is the most comprehensive set of data for the vascular networks considered.  However, it does not help to identify which subset of variables are primarily responsible for classification over other variables.  To do that, we focus our attention to the PCA results.

A common feature of vascular network data is the large variation observed in the the scaling of branch lengths \cite{s_bentley_etal_ecolett_2013, s_newberry_etal_ploscompbio_2015, s_tekin_etal_ploscompbio_2016}, often times exhibited over many orders of magnitude.  In Figure \ref{fig:tree_pca}\textbf{f} we see that the first principle component explains 34.2\% of the variance in the data, and is composed primarily of the length scale factors ($\bar{\gamma}, \Delta \gamma$).  Interestingly, the length scale factors are not powerful for classification purposes (see Table \ref{tab:features} and Figure 4 in the main text), even though they are responsible for much of the variance in the data.  On the other hand, we find that correlations between the radial scale factors account for 17.3\% and 16.4\% of the variance through the second and third principle components.  This result, in combination with the fact that the radial scale factors had the best global classification scores in Table \ref{tab:features}, led to the selection of the radial average and difference scale factors as the final choice for the feature space.

\newpage

\section{Normalizing results from classifier methods}
While the three machine learning methods tested effectively perform the same task (e.g. classification of data), the manner in which this is done varies significantly, and thus requires careful consideration when trying to compare results.  This difficulty is compounded by the inclusion of a variable classification sensitivity level.  Here we describe the process used to standardize classification output from the three methods of Kernel Density Estimation (KDE), Support Vector Machine (SVM), and Logistic Regression (LR).  Only one variable feature space was used for the comparison between methods, and that was the radial scale factor feature space ($\bar{\beta}, \Delta \beta$).  This feature space was chosen as it performed best for classification across all methods (Table \ref{tab:features}, as well as explaining up to 33\% of the variance in all of the asymmetric scale factor variables (Figure \ref{fig:tree_pca}\textbf{f-g}).  The testing groups used were that of mammal and plant.  These were chosen to increase the performance of methods reliant on dataset size for training and testing. Finally, we present receiver operating characteristic (ROC) curves that are used to perform the final comparison once standardization has been achieved.

\subsection{Kernel density estimation}
The non-parametric kernel density estimation procedure put forth by Duong et al.~\cite{s_duong_etal_pnas_2012, s_duong_jnonparstats_2013} tests for uniqueness and overlap between two different probability distributions generated from empirical data.  Probability distributions, $\mathbb{P}_i(\textbf{x, X})$ are defined on a discretized feature space, where $\textbf{x}$ represents the discretized coordinates in the feature space, $\textbf{X}$ represents actual empirically measured values, and $i = A$ or $B$ represents our known classifier.  When performing the local significant difference test, one must first set the $p$-value that denotes ``significance".  Conventionally this value is set at either $p = 0.01$ (as done for the analysis presented in the main text) or $p = 0.05$.  The KDE method then identifies regions of the feature space where the squared-difference between the probabilities, $\mathbb{P}_i(\textbf{x, X})$, is less than the selected significance level.  Finally, by examining which probability, $\mathbb{P}_i(\textbf{x, X})$, is greatest within each region, one can identify which classifier is driving differentiation.  Thus, one can measure the performance of the KDE method at a given significance level by counting the number of correctly and incorrectly classified points with respect to each region (Figure S\ref{fig:point_count_chart}).

By varying the $p$-value one can vary the relative size of the classified regions of the test to examine the efficacy of classification (Figure S\ref{fig:point_count_chart}).  For the KDE method, the $p$-value was varied by integer orders of magnitude from $10^{-12}$ to $10^4$ to ensure that the limiting scenarios of zero points classified and all points classified were contained.  Then, for each sensitivity level, correctly and incorrectly classified points were tallied for further analysis and comparison in a one-versus-all framework.

\begin{figure}[h]
\begin{centering}
\includegraphics[width = \textwidth]{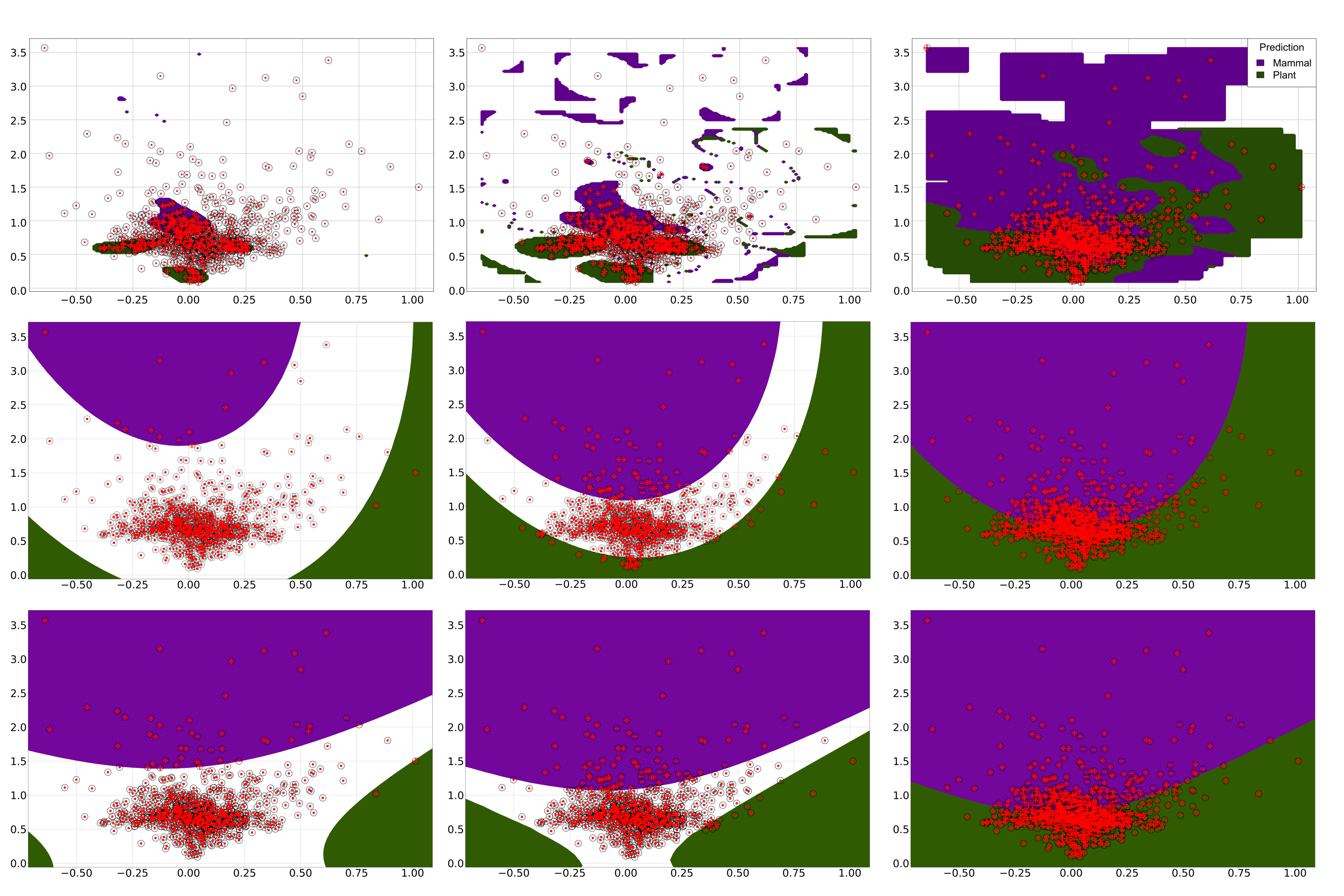}
\caption[Data classification versus sensitivity for KDE, LR, and SVM methods]{\label{fig:point_count_chart} Data classification versus significance thresholds for KDE (top row), LR (middle row), and SVM (bottom row) methods.  Significance thresholds vary from high (left column), to medium (middle column), to low (right column).  Horizontal axes all correspond to the radial difference scale factor $(\Delta \beta)$, while vertical axes all correspond to the radial average scale factor $(\bar{\beta})$.}
\end{centering}
\end{figure}

\subsection{Logistic regression and support vector machine}
The logistic regression (LR) and support vector machine (SVM) methods represent two approaches at using supervised machine learning methods for classification \cite{s_friedman_elements_statistical_learning_2016, s_aurelien_scikit_learn_2017}.  These methods differ from the KDE approach by using a training set of data to partition the feature space into two separable regions (separated by the decision boundary).  Then, points from a testing set of data are assigned a classification score based on their positions in the feature space with respect to the decision boundary.  In Figure S\ref{fig:point_count_chart} are graphs of results from the LR method and the SVM method.  In these examples, the decision boundary corresponds to a probability score of 0.5, and any testing point with a score between 0 and 0.5 is classified as a plant, while a score between 0.5 and 1 is classified as a mammal.

To compare to the output of the KDE approach, an analogue of significance to the KDE approach must be defined for the LR and SVM approaches.  In the context of the LR and SVM approaches this was done by defining significance as the distance a probability score is from 0.5.  Regions of equal significance were determined by binning along the probability score axis.  Thus, points with probability scores in the ranges of $[0, 0.05]$ for mammals or $[0.95, 1]$ for plants would all be characterized as equally, highly significant predictions (or in terms of the KDE method, would correspond to the next lowest possible $p$-value).  The bins were then successively enlarged to reflect a decrease in test significance.  So, for mammals, the bins used were $[0, 0.05]$, $[0, 0.1]$, $[0, 0.15]$, $[0, 0.2]$, $[0, 0.25]$, $[0, 0.3]$, $[0, 0.35]$, $[0, 0.4]$, $[0, 0.45]$, $[0, 0.5]$.  For plants, the bins used were $[0.95, 1]$, $[0.9, 1]$, $[0.85, 1]$, $[0.8, 1]$, $[0.75, 1]$, $[0.7, 1]$, $[0.65, 1]$, $[0.6, 1]$, $[0.55, 1]$, $[0.5, 1]$.  In Figure S\ref{fig:point_count_chart} are graphs of results from the LR and SVM methods showing the successive binning approach.  Finally, for each binned region corresponding to varying levels of significance, correctly and incorrectly classified points were identified for comparison in a one-versus-all framework.

\subsection{ROC comparison of methods}
To finally compare the three methods of kernel density estimation (KDE), logistic regression (LR), and support vector machine (SVM), we graphed receiver operating characteristic (ROC) curves of the results of the methods as sensitivities were varied \cite{s_friedman_elements_statistical_learning_2016, s_aurelien_scikit_learn_2017}.  ROC curves are graphs of a methods true positive rate (TPR) versus its false positive rate (FPR).  
Due to the manner in which we are comparing machine learning methods, we technically have three classes for any given significance level: mammals, plants, and undetected.  Thus, we must use a one-versus-all framework for calculating true and false positives.  In calculating the TPR and FPR, we first choose which category we are ``detecting'', say plants, then the TPR and FPR can be calculated as,

\begin{align}
TPR & = \frac{N_{plant +}}{N_{plant +} + N_{not~plant -}} \nonumber \\ 
\label{eq:plant_positive} \\
FPR & = \frac{N_{plant -}}{N_{plant -} + N_{not~plant +}} \nonumber
\end{align}

where $N_{plant \pm}$ corresponds to the number of data points correctly (or incorrectly) classified as plant as denoted by the $+$ sign ($-$ sign) within the green plant contours in Figure \ref{fig:machine}, and $N_{not~plant \pm}$ corresponds to the number of data points correctly (or incorrectly) classified as not plant external to the green plant contours.  A TPR and an FPR is then calculated for each level of significance (defined by a $p$-value for the KDE approach, or a probability bin for the LR or SVM approaches).  Finally, the ROC curve can be graphed, as presented in Figure \ref{fig:machine}\textbf{c}.

Conventionally, when comparing classification schemes using an ROC curve, the best method is identified as whichever method sits most in the upper-left corner of the graph, or which ever has the greatest ``area under the curve'' (AUC), as this represents a maximal true positive rate and a minimal false positive rate.  To use the ROC curves for the different methods of KDE, LR, and SVM, we can use Eq.~(\ref{eq:plant_positive}) and its mammal classifying analogue, resulting in Figure \ref{fig:machine}\textbf{c}.  Here we can observe that overall the KDE method outperforms the LR and SVM methods as having either a lower FPR for a given TPR (plant as positive class graph on left in Figure \ref{fig:machine}\textbf{c}) or a greater TPR for a given FPR (mammal as positive class graph on right in Figure \ref{fig:machine}\textbf{c}).

\subsection{Data grouping}
Once selection of the classifier was made, the subgroups of datasets were prepared for higher resolution classification.  Using the KDE method, individual HHT and ponderosas were found indistinguishable, as were species within the GS Tips and AS Tips groups, hence their final merging into larger datasets (see Tables S4-6).  We obtained 8 different major species/groups: HHT, mouse lung, Balsa, Pi\~{n}on, Ponderosa, GS Tips, AS Tips, and Roots, each with 6 recorded variables: $\bar{\beta}, \bar{\gamma}, \Delta\beta, \Delta\gamma,$ and the merging of $\beta_1$ and $\beta_2$ into a single distribution as well as for $\gamma_1$ and $\gamma_2$.  Specific groupings of the scale factor variables used were: ($\beta_1, \beta_2; \gamma_1, \gamma_2$) as a two-dimensional distribution representative of the \textit{symmetric} formalism; ($\bar{\beta}, \bar{\gamma}, \Delta\beta, \Delta\gamma$) as a four-dimensional distribution representative of the \textit{asymmetric} formalism; ($\bar{\beta}, \bar{\gamma}$) as a two-dimensional distribution for \textit{average} scaling; ($\Delta\beta, \Delta\gamma$) as a two-dimensional distribution for \textit{difference} scaling; ($\bar{\beta}, \Delta\beta$) as a two-dimensional distribution for radial scaling; and ($\bar{\gamma}, \Delta\gamma$) as a two-dimensional distribution for length scaling.  For the full list of comparison results using the KDE method, see Tables S1-3.

\newpage

\section{Derivation of exact metabolic scaling exponent formula}

Here we present a derivation of the metabolic scaling exponent under the general assumptions of the West, Brown, Enquist model for vascular branching \cite{s_west_etal_science_1997, s_savage_etal_ploscompbio_2008, s_brummer_etal_ploscompbio_2017}.  This approach deviates from conventional derivations in that zero approximations will be made regarding network size or the particulars of volume scaling outside of the bounds of self-similarity.  Some discussion will be presented at the end regarding how network geometry can be included in terms of symmetric or asymmetric branching.

The typical starting point for modeling metabolic scaling is Kleiber's Law \cite{kleiber_hilgardia_1932}, the empirically motivated, power-law relationship between organismal basal metabolic rate and mass, expressed as,

\begin{equation}
B = B_0\left(\frac{M}{M_0}\right)^\theta
\end{equation}

Here $B$ represents the measured metabolic rate, $M$ the mass, $\theta$ the metabolic scaling exponent observed to cluster around 3/4 \cite{s_kleiber_hilgardia_1932, s_savage_etal_funceco_2004}, and $B_0$ and $M_0$ normalization constants.  Treating metabolic rate as the combined sum of metabolism of every terminal branch in an organism, we can substitute $B = B_\TIP N_\TIPS$, where $B_\TIP$ represents total metabolism per terminal branch and $N_\TIPS$ the total number of metabolizing terminal branches.  Doing so results in,

\begin{equation}
B_\TIP N_\TIPS = B_0\left( \frac{M}{M_0}\right)^\theta
\end{equation}

Next, mass is substituted with total volume.  The validity of this substitution is the result of having optimized the geometrical scaling of a hierarchically branching vascular network against the dual demands of hydrodynamic resistance to fluid flow and fractal space-filling \cite{s_west_etal_science_1997, s_savage_etal_ploscompbio_2008, s_brummer_etal_ploscompbio_2017}.  Performing the substitution gives us,

\begin{equation}
N_\TIPS = \left( \frac{V_\TOT}{V_0} \right)^\theta
\label{eq:regression}
\end{equation}

where we have cancelled out $B_\TIP$ with $B_0$.  At this point it is important to pause and recognize Eq.~\ref{eq:regression} as a method by which one can estimate the metabolic scaling exponent in a vascular organism free from explicitly imposing assumptions regarding network geometry via symmetric or asymmetric branching.  One powerful aspect of Eq.~\ref{eq:regression} is that it can be applied recursively on any individual vascular branching network.  Specifically, for every branch (or vessel) in a network, the total number of distal terminal branches can be represented by $N_\TIPS$, and the total volume of all distal branches are represented by $V_\TOT$.  Thus, a standard major axis regression analysis can be performed to determine the value of $\theta$ that corresponds to an individual organism as per this model.  An example of such an analysis being performed on the mouse lung data set is presented in Figure S\ref{fig:ml_regression}.

\begin{figure}[h]
\begin{centering}
\includegraphics[width = 0.75\textwidth]{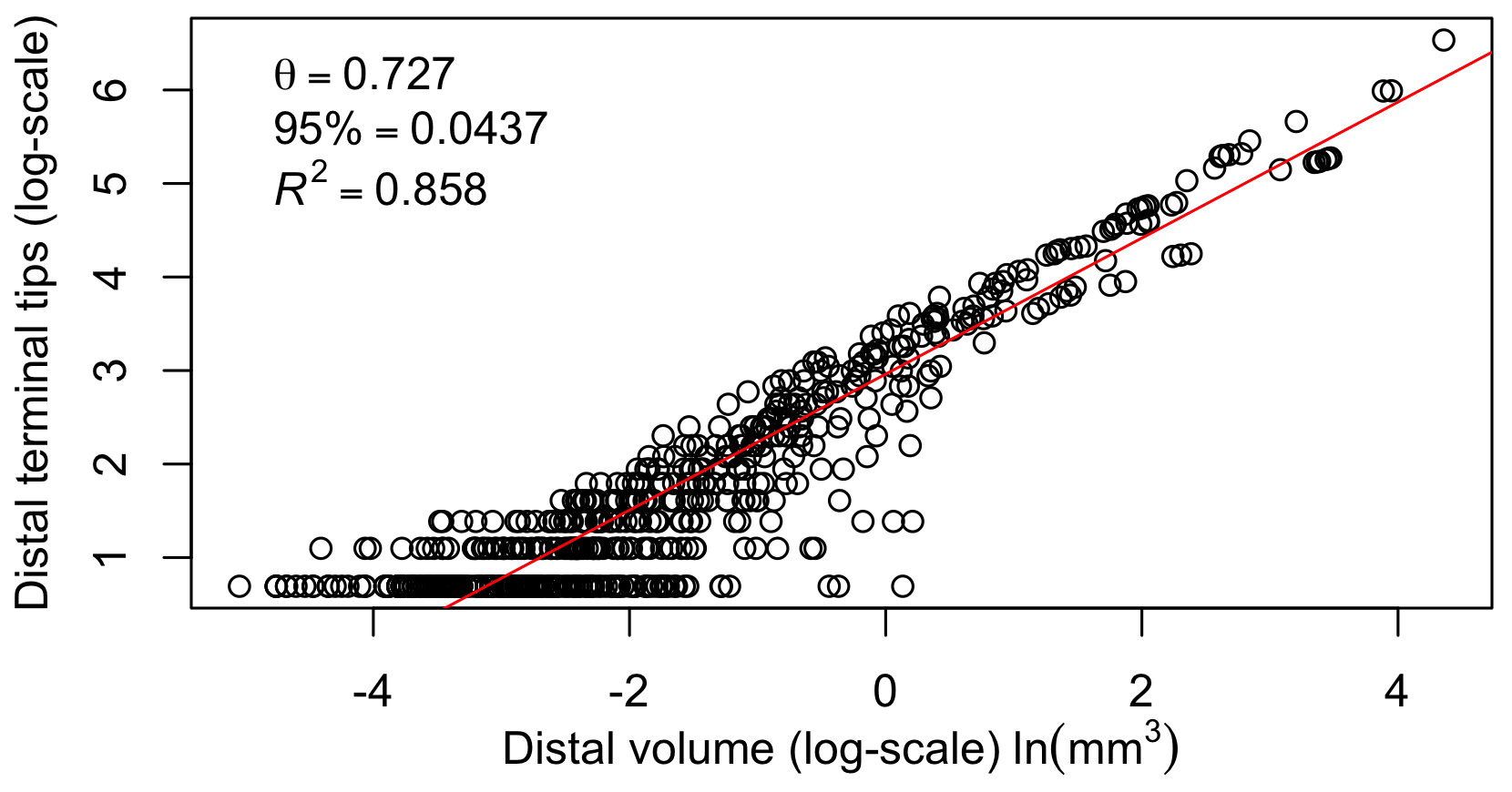}
\caption[Sample regression analysis on mouse lung data]
{\label{fig:ml_regression} Sample regression analysis of distal terminal tips to distal vessel volume on mouse lung data using Standard Major Axis (SMA) regression.  The fit line corresponding to the reported value of $\theta$ is represented in red.}
\end{centering}
\end{figure}

It is at this point that we must specify in greater detail the functional form of the total volume of the vascular network, $V_\TOT$.  Assuming that we are working with a strictly self-similar, hierarchically branching, pipe-like model, then the total volume of the network can be expressed as

\begin{equation}
V_\TOT = V_{\N,\TOT} \sum_{j = 0}^N \nu^{-j}
\label{eq:v_tot_sum}
\end{equation}

where $j$ and $N$ represent the $j^{th}$ and $N^{th}$ generations of the network, $V_{\N,\TOT}$ is the total volume of the terminal ($N^{th}$) generation, and $\nu$ represents the ratio of total volume from sibling branches to their respective parent branch.  For example, in a bifurcating symmetric network, $\nu = 2(\pi r_{j+1}^2 l_{j+1})/(\pi r_j^2l_j) = 2\beta^2 \gamma$.  Recognizing Eq.~\ref{eq:v_tot_sum} as a geometric series, we can write it's exact form as,

\begin{equation}
V_\TOT = V_{\N, \TOT} \frac{1 - \nu^{-(N+1)}}{1 - \nu^{-1}}
\end{equation}

which is valid for all values of $\nu$ except for $\nu \approx  1$.  It is not uncommon to find individual organisms with $\nu \approx 1$.  This scenario is problematic if using the above formula due to its asymptotic nature.  However, this problem can be remedied by using L'H\^{o}pital's rule, producing the piecewise result,

\begin{equation}
V_{\TOT} = 
\begin{dcases} 
	V_{\N,\TOT} \frac{1 - \nu^{-(N+1)}}{1 - \nu^{-1}} & \mbox{for } \nu \ne 1 \\
	V_{\N,\TOT}  \frac{N+1}{\nu^N} & \mbox{for } \nu \approx 1 
\end{dcases}
\label{eq:piecewise_volume}
\end{equation}
\\

Upon substituting Eq.~\ref{eq:piecewise_volume} into Eq.~\ref{eq:regression}, we arrive at the following exact, piecewise function for the metabolic scaling exponent,

\begin{equation}
\theta = 
\begin{dcases} 
	\frac{\ln(2^N)}{\ln(2^N) + \ln(1 - \nu^{N+1}) - \ln(\nu^N(1-\nu))} & \mbox{for } \nu \ne 1 \\
	\frac{\ln(2^N)}{\ln((N+1)2^N) - N\ln(\nu)} & \mbox{for } \nu \approx 1 
\end{dcases}
\label{eq:s_mse_exact}
\end{equation}
\\

where we set $V_{\N, \TOT} = N_{\TIPS}V_{\TIP}$, $V_0$ was cancelled out with $V_{\TIP}$, and we have restricted ourselves to considering strictly bifurcating networks such that $N_{\TIPS} = 2^{N}$.  One benefit of Eq.~\ref{eq:s_mse_exact} is that we can easily examine the functional relationship between metabolic rate and the scaling of volume in a general sense, as presented in Figure S\ref{fig:piecewise_scaling}.  Examining the influence of symmetric or asymmetric branching can then be done separately.

\begin{figure}[h]
\begin{centering}
\includegraphics[width = 0.75\textwidth]{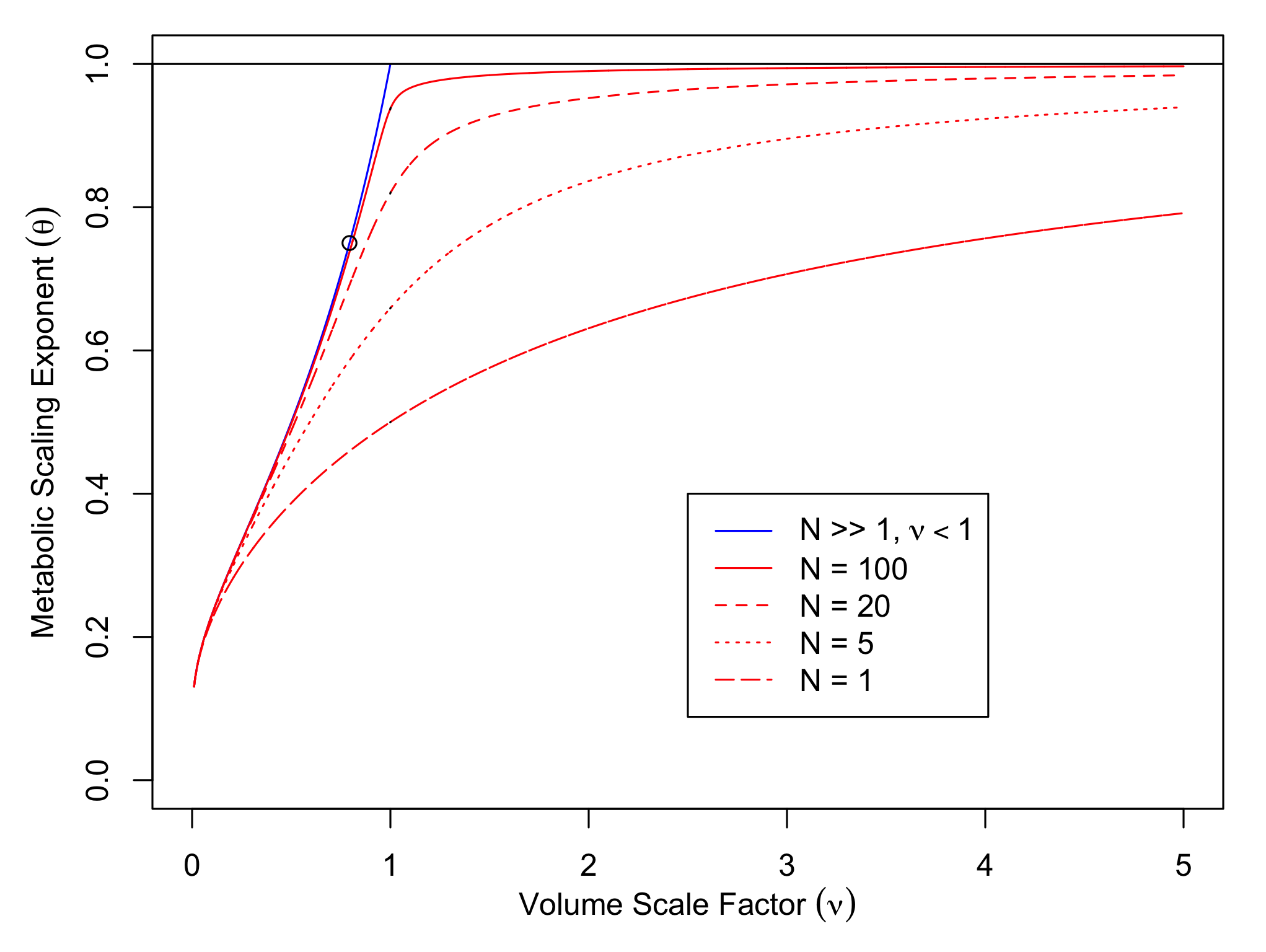}
\caption[Exact formula for metabolic scaling exponent versus volume scaling and network size]
{\label{fig:piecewise_scaling} Exact formula for metabolic scaling exponent versus volume scaling and network size.  In red are presented different exact metabolic scaling exponent curves corresponding to different values of total network generations $N$.  In blue is the curve for the large network size (large $N$) and sub-linear volume scaling ($\nu < 1$) approximation.  The black circle represents the particular scenario of 3/4 metabolic scaling corresponding to the symmetric WBE predicted value for $\nu = n \beta^2\gamma = 1/2^{1/3}$.}
\end{centering}
\end{figure}

Several important features are present in Figure S\ref{fig:piecewise_scaling}.  Most notably is the effect that network size, $N$, has on varying the sensitivity of the metabolic scaling exponent to the value of the volume scale factor, $\nu$.  For all values of network size, the metabolic scaling exponent is bounded between zero and one.  For the case of large networks ($N = 100$ in Figure S\ref{fig:piecewise_scaling}), the metabolic scaling exponent nearly reaches the asymptotic value of $\theta = 1$ when $\nu = 1$, where as for smaller size networks (lesser values of $N$), the rate of increase in $\theta$ is markedly less.  This result lends support to the argument that the approximate form of Eq.~\ref{eq:s_mse_exact}, namely $\theta \approx \ln(2)/(\ln(2) - \ln(2\beta^2\gamma))$ for a symmetrically branching network, holds in the limit that $N >> 1$ and $\nu < 1$.  This approximate form of $\theta$ is presented in Figure S\ref{fig:piecewise_scaling} by the solid, blue line.

A second important result present in Figure S\ref{fig:piecewise_scaling} is the fact that all network sizes can take on the empirically observed value of $\theta = 3/4$ as long as $\nu$ is large enough.  This result supports previous arguments regarding transitions in blood flow related to shifts in the scaling of radii.  In these circumstances, the radial scaling transitions from cross-sectional area-preserving ($\beta^2 = 1/2$ for a symmetrically bifurcating network) to area-increasing ($\beta^3 = 1/2$).  Due to the conservation fluid flow-rates across branching junctions, this latter scaling acts to slow the flow of blood.  However, a simultaneous result of changing the scaling of radii in such a way is to increase the volume scale factor, assuming that there is no change in the scaling of lengths.  Thus, this predicted trade-off between decreasing network size and increasing volume scaling would appear consistent with the biological demands associated with the cardiovascular system in mammals.  As for plants, ontogenetic shifts in metabolic scaling (as measured through total respiration) have been reported between sapling and mature trees and shrubs, yet not in the context of the measurement of vascular branching traits \cite{s_mori_etal_pnas_2010}.  Thus, it would be interesting to see if future studies corroborate trade-offs between metabolic scaling exponents, vascular branching traits, and network size.

In using Eq.~\ref{eq:s_mse_exact}, one must specify (and thereby either measure or estimate) the total number of branching generations in a vascular network. Unfortunately this is not a particularly straightforward task.  With the conventional WBE generational labeling scheme, the generation index is advanced once a branching occurs for all branches within the previous generation.  For highly asymmetric networks this can result in having branches of wildly differing size be members of the same generation.  Alternative labelling schemes have been proposed outside the context of the WBE framework, specifically that of Horton-Strahler \cite{s_horton_geosocbull_1945, s_strahler_eos_1957}.  While capable of handling moderate levels of asymmetry, one can show that the Horton-Strahler labelling scheme is insufficient at handling networks resembling the fishbone branching pattern.  For the purposes of this work, we adopt an approach based on the asymptotically symmetric expectation that for a network with $N$ generations, there should be approximately $2^N$ terminal branches.  Thus, from a count of the  number of terminal branches, $N_{\TIPS}$, one can estimate the number of generations as $N = \ln(N_{\TIPS})/\ln(2)$.

Lastly, in the main texts standard errors are reported for calculated values of the metabolic scaling exponent in Figure 4\textbf{a}.  These errors were determined from standard methods, namely $\sigma_\theta^2 \approx \left\|\frac{\partial \theta}{\partial \nu}\right\|^2\sigma_\nu$, where $\sigma_\nu$ is determined from the chosen symmetric or asymmetric scale factor variables.


\newpage

\section{Derivation of curvature in metabolic rate versus mass}

Here we present a derivation of the curvature in the metabolic rate, $B$, of an organism in terms versus its mass, in log-log space.  Beginning again with Kleiber's Law after having linearized the equation, we have,

\begin{equation}
\ln(B) = \ln(B_0) + \theta \ln \left( \frac{M}{M_0} \right)
\end{equation}

As we are interested in examining the mass related curvature in metabolic rate, we must evaluate 
$\partial^2 \ln(B)/\partial (\ln(M))^2$.  Since $\theta$ is effectively a mass-dependent quantity through its dependence on network size via total generation number $N$, we need to find an appropriate substitute for mass.  Assuming that the total mass of a vascular organism can be expressed asymptotically as the sum total of all cells serviced by the vascular architecture, then $M \approx M_{\TIP}2^N$.  Thus, we can build a derivative operator as,

\begin{equation}
\frac{\partial}{\partial \ln(M)} = \frac{1}{\ln(2)} \frac{\partial}{\partial N}
\end{equation}

Using this operator, curvature in metabolic rate can be expressed as follows,

\begin{equation}
\frac{\partial^2 \ln(B)}{\partial (\ln(M))^2} = \frac{2}{\ln(2)} \frac{\partial \theta}{\partial N} + \frac{N}{\ln(2)} \frac{\partial^2 \theta}{\partial N^2}
\label{eq:curvature_compact}
\end{equation}

where we have maintained the explicit notation for mass, $M$, on the left hand side as a reminder of what the quantity on the right represents.  Now the problem is reduced to calculating derivatives of the metabolic scaling exponent $\theta$ with respect to total number of generations $N$.  An important and immediate consequence of Eq.~\ref{eq:curvature_compact} is that the approximate ($\nu < 1$) and asymptotic ($N >> 1$) version of Eq.~(\ref{eq:s_mse_exact}), $\theta \approx \ln(2)/\left[ \ln(2) - \ln(\nu) \right]$, results in zero curvature due to its invariance in relative network size.  This prediction is consistent with alternative parameterizations of curvature in metabolic scaling \cite{s_kolokotrones_etal_nature_2010}, and could inform observed deviations from previous theoretical predictions of metabolic scaling exponents as discussed in the main text.

Focusing on the version of Eq.~(\ref{eq:s_mse_exact}) for $\nu \ne 1$, we find that the curvature in $B$ can be expressed finally in the relatively compact form of,

\begin{equation}
\frac{\partial^2 \ln(B)}{\partial (\ln(M))^2} = \frac{1}{x^4} \left \{ 2x^3 + 2N^2x\left( \frac{\partial x}{\partial N} \right)^2 - 4Nx^2 \frac{\partial x}{\partial N} - N^2x^2 \frac{\partial^2x}{\partial N^2} \right \}
\label{eq:curvature_exact}
\end{equation}

where $x = \ln \left(V_{\TOT}/V_{\TIP} \right)$.  Graphs of Eq.~\ref{eq:curvature_exact} are presented in the main text, where we find that curvature is predicted to always be positive, consistent with previous studies based on mammalian respiration \cite{s_kolokotrones_etal_nature_2010}, but inconsistent with studies based on plant respiration \cite{s_mori_etal_pnas_2010}.  Noting the form of $x$, we can identify that $x \propto N$, thus curvature is proportional to $1/N$ and decreases with an increasing number of generations.  This network size-based suppression of curvature has been reported in relation to plant respiration \cite{s_mori_etal_pnas_2010}.

These results demonstrate a current knowledge gap within the field of metabolic scaling, and motivate the simultaneously conducting respiration-based measurements of metabolic rates and vascular imaging to better synchronize prediction with observation.

\newpage

\section{Kernel Density Estimation Results}

In this section we present the full results from the kernel density estimation procedure spanning all combinations classification features related to geometric scaling variables.

\captionsetup[table]{labelfont={bf}, labelformat={default}, labelsep=space, name={Table S\hspace{-1mm}}}
\begin{table}
\caption[Table of \textit{p}-values for KDE test on plant animal groups I.]
{\label{tab:group_pval}\textbf{$|$  Results of pairwise classification of plants and animal groups using asymmetric scale factors related to hydraulics and fluid transport ($\bar{\beta}, \Delta \beta$) or the strictly symmetric branching scale factors ($\beta, \gamma$)}}
\resizebox{\textwidth}{!}{
\begin{tabular}{|l|c|c|c|c|c|c|c|c|}
  \hline
  \rowcolor{gray!50}
  \backslashbox{SYM}{ASYM} & HHT* & Mouse Lung & Balsa & Pi\~{n}on & Ponderosa* & GS Tips*& AS Tips* & Roots \\ 
  \hline
	\cellcolor{gray!50}HHT* & $-$ & \cellcolor{gray!25} $2\times10^{-1}$ & \cellcolor{gray!25} $6\times10^{-13}$ & \cellcolor{gray!25} $1\times10^{-4}$ & \cellcolor{gray!25} $3\times10^{-2}$ & \cellcolor{gray!25} $9\times10^{-9}$ & \cellcolor{gray!25} $1\times10^{-1}$ & \cellcolor{gray!25} $2\times10^{-2}$ \\ 
  \hline
	\cellcolor{gray!50}Mouse Lung & $6\times10^{-2}$ & $-$ & \cellcolor{gray!25} $5\times10^{-18}$ & \cellcolor{gray!25} $7\times10^{-3}$ & \cellcolor{gray!25} $7\times10^{-5}$ & \cellcolor{gray!25} $2\times10^{-19}$ & \cellcolor{gray!25} $2\times10^{-2}$ & \cellcolor{gray!25} $2\times10^{-6}$ \\ 
  \hline
	\cellcolor{gray!50}Balsa & $1\times10^{-1}$ & $4\times10^{-4}$ & $-$ & \cellcolor{gray!25} $1\times10^{-16}$ & \cellcolor{gray!25} $6\times10^{-2}$ & \cellcolor{gray!25} $1\times10^{-6}$ & \cellcolor{gray!25} $6\times10^{-2}$ & \cellcolor{gray!25} $6\times10^{-13}$ \\ 
  \hline
	\cellcolor{gray!50}Pi\~{n}on & $3\times10^{-7}$ & $3\times10^{-6}$ & $2\times10^{-7}$ & $-$ & \cellcolor{gray!25} $4\times10^{-13}$ & \cellcolor{gray!25} $3\times10^{-56}$ & \cellcolor{gray!25} $2\times10^{-5}$ & \cellcolor{gray!25} $2\times10^{-13}$ \\ 
  \hline
	\cellcolor{gray!50}Ponderosa* & $2\times10^{-1}$ & $7\times10^{-2}$ & $2\times10^{-3}$ & $4\times10^{-4}$ & $-$ & \cellcolor{gray!25} $9\times10^{-5}$ & \cellcolor{gray!25} $3\times10^{-1}$ & \cellcolor{gray!25} $1\times10^{-4}$\\ 
  \hline
	\cellcolor{gray!50}GS Tips* & $9\times10^{-3}$ & $3\times10^{-2}$ & $8\times10^{-4}$ & $1\times10^{-3}$ & $1\times10^{-1}$ & $-$ & \cellcolor{gray!25} $7\times10^{-12}$ & \cellcolor{gray!25} $2\times10^{-11}$ \\ 
  \hline
	\cellcolor{gray!50}AS Tips* & $2\times10^{-1}$ & $1\times10^{-1}$ & $7\times10^{-3}$ & $2\times10^{-2}$ & $4\times10^{-1}$ & $3\times10^{-2}$ & $-$ & \cellcolor{gray!25} $5\times10^{-3}$ \\ 
  \hline
	\cellcolor{gray!50}Roots & $1\times10^{-1}$ & $6\times10^{-3}$ & $2\times10^{-3}$ & $2\times10^{-5}$ & $1\times10^{-1}$ & $3\times10^{-2}$ & $3\times10^{-1}$ & $-$ \\ 
   \hline
\end{tabular}
}
\caption*{Table of \textit{p}-values for pairwise KDE testing on plant and animal groups comparing different feature spaces.  The lower diagonal corresponds to scale factors associated with strictly symmetric branching $(\beta, \gamma)$, while the upper diagonal corresponds to scale factors associated with hydraulics and asymmetric radial branching $(\bar{\beta}, \Delta \beta)$. Significant classification is defined as occurring when \textit{p}$ < 0.05$.  Datasets with multiple individuals that were aggregated are indicated with asterisks (*).  Of note is the universally enhanced ability of the radial scale factor feature space to differentiate between groups when compared to the symmetric scale factor feature space.}
\end{table}

\captionsetup[table]{labelfont={bf}, labelformat={default}, labelsep=space, name={Table S\hspace{-1mm}}}
\begin{table}
\caption[Table of \textit{p}-values for KDE test on plant animal groups II.]
{\label{tab:all_length_pval}\textbf{$|$ Results of pairwise classification of plant and animal groups using asymmetric scale factors that are related to space-filling only ($\bar{\gamma}, \Delta \gamma$) or space-filling and hydraulics ($\bar{\beta}, \Delta \beta, \bar{\gamma}, \Delta \gamma$).}}
\resizebox{\textwidth}{!}{
\begin{tabular}{|l|c|c|c|c|c|c|c|c|}
  \hline
  \rowcolor{gray!50}
  \backslashbox{ALL}{LENGTH} & HHT* & Mouse Lung & Balsa & Pi\~{n}on & Ponderosa* & GS Tips*& AS Tips* & Roots \\ 
	\hline
	\cellcolor{gray!50} HHT* & $-$ & \cellcolor{gray!25} $7 \times 10^{-2}$ & \cellcolor{gray!25} $8 \times 10^{-1}$ & \cellcolor{gray!25} $6 \times 10^{-3}$ & \cellcolor{gray!25} $1 \times 10^{-1}$ & \cellcolor{gray!25} $1 \times 10^{-2}$ & \cellcolor{gray!25} $3 \times 10^{-2}$ & \cellcolor{gray!25} $8 \times 10^{-3}$ \\ 
	\hline
	\cellcolor{gray!50} Mouse Lung & $1 \times 10^{-1}$ & $-$ & \cellcolor{gray!25} $1 \times 10^{-1}$ & \cellcolor{gray!25} $2 \times 10^{-3}$ & \cellcolor{gray!25} $4 \times 10^{-1}$ & \cellcolor{gray!25} $4 \times 10^{-1}$ & \cellcolor{gray!25} $9 \times 10^{-2}$ & \cellcolor{gray!25} $9 \times 10^{-6}$ \\ 
	\hline
	\cellcolor{gray!50} Balsa & $9 \times 10^{-5}$ & $2 \times 10^{-5}$ & $-$ & \cellcolor{gray!25} $2 \times 10^{-2}$ & \cellcolor{gray!25} $1 \times 10^{-2}$ & \cellcolor{gray!25} $1 \times 10^{-3}$ & \cellcolor{gray!25} $1 \times 10^{-2}$ & \cellcolor{gray!25} $3 \times 10^{-2}$ \\ 
	\hline
	\cellcolor{gray!50} Pi\~{n}on & $4 \times 10^{-11}$ & $4 \times 10^{-8}$ & $1 \times 10^{-8}$ & $-$ & \cellcolor{gray!25} $3 \times 10^{-1}$ & \cellcolor{gray!25} $6 \times 10^{-2}$ & \cellcolor{gray!25} $5 \times 10^{-1}$ & \cellcolor{gray!25} $3 \times 10^{-1}$ \\ 
	\hline
	\cellcolor{gray!50} Ponderosa* & $8 \times 10^{-3}$ & $9 \times 10^{-3}$ & $7 \times 10^{-3}$ & $6 \times 10^{-5}$ & $-$ & \cellcolor{gray!25} $3 \times 10^{-1}$ & \cellcolor{gray!25} $5 \times 10^{-1}$ & \cellcolor{gray!25} $1 \times 10^{-1}$ \\ 
	\hline
	\cellcolor{gray!50} GS Tips* & $2 \times 10^{-4}$ & $6 \times 10^{-4}$ & $1 \times 10^{-3}$ & $1 \times 10^{-5}$ & $2 \times 10^{-1}$ & $-$ & \cellcolor{gray!25} $3 \times 10^{-1}$ & \cellcolor{gray!25} $9 \times 10^{-4}$ \\ 
	\hline
	\cellcolor{gray!50} AS Tips* & $5 \times 10^{-2}$ & $2 \times 10^{-2}$ & $1 \times 10^{-1}$ & $1 \times 10^{-2}$ & $4 \times 10^{-1}$ & $1 \times 10^{-1}$ & $-$ & \cellcolor{gray!25} $3 \times 10^{-1}$ \\ 
	\hline
	\cellcolor{gray!50} Roots & $3 \times 10^{-4}$ & $1 \times 10^{-4}$ & $8 \times 10^{-8}$ & $1 \times 10^{-3}$ & $2 \times 10^{-2}$ & $1 \times 10^{-3}$ & $4 \times 10^{-2}$ & $-$ \\ 
   \hline
\end{tabular}
}
\caption*{Table of \textit{p}-values for pairwise KDE testing on plant and animal groups comparing different features.  The lower diagonal corresponds to scale factors associated with asymmetric branching related to space-filling ($\bar{\gamma}, \Delta \gamma$) while the upper diagonal corresponds to scale factors associated with the combined asymmetric feature spaces related to both hydraulics and space-filling ($\bar{\beta}, \Delta \beta, \bar{\gamma}, \Delta \gamma$).  Significant classification is defined as occurring when \textit{p}$ < 0.05$.  Datasets with multiple individuals that were aggregated are indicated with asterisks (*).  In this scenario we see that nearly all groups (excluding the AS Tips) are more differentiable when using all scale factors for the feature space when compared to the length scale factor feature space.}
\end{table}

\captionsetup[table]{labelfont={bf}, labelformat={default}, labelsep=space, name={Table S\hspace{-1mm}}}
\begin{table}
\caption[Table of \textit{p}-values for KDE test on plant animal groups III.]
{\label{tab:ave_diff_pval}\textbf{$|$ Results of pairwise classification of plant and animal groups using asymmetric scale factors that are related to hydraulics and space filling and that are associated with the central means of asymmetric branching ($\bar {\beta}, \bar{\gamma}$) or scale factors associated with variation in asymmetric branching ($\Delta \beta, \Delta \gamma$).}}
\resizebox{\textwidth}{!}{
\begin{tabular}{|l|c|c|c|c|c|c|c|c|}
  \hline
  \rowcolor{gray!50}
  \backslashbox{AVE}{DIFF} & HHT* & Mouse Lung & Balsa & Pi\~{n}on & Ponderosa* & GS Tips*& AS Tips* & Roots \\ 
  	\hline
	\cellcolor{gray!50}HHT* & $-$ & \cellcolor{gray!25} $4 \times 10^{-1}$ & \cellcolor{gray!25} $4 \times 10^{-3}$ & \cellcolor{gray!25} $5 \times 10^{-3}$ & \cellcolor{gray!25} $3 \times 10^{-1}$ & \cellcolor{gray!25} $3 \times 10^{-1}$ & \cellcolor{gray!25} $2 \times 10^{-1}$ & \cellcolor{gray!25} $4 \times 10^{-2}$ \\
  	\hline
  	\cellcolor{gray!50}Mouse Lung & $2 \times 10^{-1}$ & $-$ & \cellcolor{gray!25} $3 \times 10^{-2}$ & \cellcolor{gray!25} $1 \times 10^{-3}$ & \cellcolor{gray!25} $3 \times 10^{-1}$ & \cellcolor{gray!25} $3 \times 10^{-1}$ & \cellcolor{gray!25} $3 \times 10^{-1}$ & \cellcolor{gray!25} $7 \times 10^{-3}$ \\ 
	\hline
	\cellcolor{gray!50}Balsa & $3 \times 10^{-2}$ & $8 \times 10^{-7}$ & $-$ & \cellcolor{gray!25} $2 \times 10^{-4}$ & \cellcolor{gray!25} $2 \times 10^{-1}$ & \cellcolor{gray!25} $8 \times 10^{-2}$ & \cellcolor{gray!25} $7 \times 10^{-2}$ & \cellcolor{gray!25} $1 \times 10^{-5}$ \\ 
	\hline
	\cellcolor{gray!50}Pi\~{n}on & $4 \times 10^{-2}$ & $1 \times 10^{-2}$ & $3 \times 10^{-6}$ & $-$ & \cellcolor{gray!25} $1 \times 10^{-1}$ & \cellcolor{gray!25} $8 \times 10^{-2}$ & \cellcolor{gray!25} $2 \times 10^{-1}$ & \cellcolor{gray!25} $8 \times 10^{-2}$ \\ 
	\hline
\cellcolor{gray!50}Ponderosa* & $3 \times 10^{-2}$ & $1 \times 10^{-3}$ & $3 \times 10^{-2}$ & $1 \times 10^{-4}$ & $-$ & \cellcolor{gray!25} $6 \times 10^{-1}$ & \cellcolor{gray!25} $5 \times 10^{-1}$ & \cellcolor{gray!25} $9 \times 10^{-2}$ \\ 
	\hline
\cellcolor{gray!50}GS Tips* & $3 \times 10^{-6}$ & $1\times 10^{-11}$ & $3 \times 10^{-7}$ & $2 \times 10^{-12}$ & $9 \times 10^{-2}$ & $-$ & \cellcolor{gray!25} $5 \times 10^{-1}$ & \cellcolor{gray!25} $3 \times 10^{-2}$ \\ 
	\hline
\cellcolor{gray!50}AS Tips* & $1 \times 10^{-1}$ & $1 \times 10^{-2}$ & $9 \times 10^{-2}$ & $6 \times 10^{-2}$ & $3 \times 10^{-1}$ & $4 \times 10^{-3}$ & $-$ & \cellcolor{gray!25} $1\times 10^{-1}$ \\
	\hline 
\cellcolor{gray!50}Roots & $1 \times 10^{-1}$ & $4 \times 10^{-3}$ & $2 \times 10^{-2}$ & $1 \times 10^{-1}$ & $9 \times 10^{-2}$ & $1 \times 10^{-3}$ & $4 \times 10^{-1}$ & $-$ \\ 
   \hline
\end{tabular}
}
\caption*{Table of \textit{p}-values for pairwise KDE testing on plant and animal groups comparing feature spaces. The lower diagonal corresponds to scale factors associated with the central means of asymmetric branching ($\bar {\beta}, \bar{\gamma}$) while the upper diagonal corresponds to scale factors associated with variation in asymmetric branching ($\Delta \beta, \Delta \gamma$).  Significant classification is defined as occurring when \textit{p}$ < 0.05$.  Datasets with multiple individuals that are aggregated are indicated with asterisks (*).  Of note is the enhanced differentiation that occurs when using the average scale factors as the feature space--in particular for the Mouse lung, Balsa, Pi\~{n}on, Ponderosas, and GS Tips.  Interestingly, the Roots exhibit greater differentiation between all other groups using the difference scale factors as the feature space.}

\end{table}

\captionsetup[table]{labelfont={bf}, labelformat={default}, labelsep=space, name={Table S\hspace{-1mm}}}
\begin{table}
\caption[Table of \textit{p}-values for KDE test on Ponderosa individuals.]
{\label{tab:pond_pval}\textbf{$|$ Results of pairwise classification of Ponderosa individuals using scale factors that are related to hydraulics, and fluid transport ($\bar{\beta}, \Delta \beta$) and that are associated with strictly symmetric branching ($\beta, \gamma$) or asymmetric branching.}}
\resizebox{\textwidth}{!}{
\begin{tabular}{|l|c|c|c|c|c|}
  \hline
  \rowcolor{gray!50}
  \backslashbox{SYM}{ASYM} & POND03 & POND05 & POND06 & POND07 & POND16 \\ 
  \hline
	\cellcolor{gray!50}POND03 & $-$ & \cellcolor{gray!25} $5 \times 10^{-2}$ & \cellcolor{gray!25} $5 \times 10^{-1}$ & \cellcolor{gray!25} $7 \times 10^{-1}$ & \cellcolor{gray!25} $3 \times 10^{-1}$ \\ 
	\hline
	\cellcolor{gray!50}POND05 & $5 \times 10^{-1}$ & $-$ & \cellcolor{gray!25} $3 \times 10^{-2}$ & \cellcolor{gray!25} $8 \times 10^{-2}$ & \cellcolor{gray!25} $3 \times 10^{-1}$ \\ 
	\hline
	\cellcolor{gray!50}POND06 & $3 \times 10^{-1}$ & $4 \times 10^{-1}$ & $-$ & \cellcolor{gray!25} $4 \times 10^{-1}$ & \cellcolor{gray!25} $3 \times 10^{-1}$ \\ 
	\hline
	\cellcolor{gray!50}POND07 & $6 \times 10^{-1}$ & $4 \times 10^{-1}$ & $5 \times 10^{-1}$ & $-$ & \cellcolor{gray!25} $6 \times 10^{-1}$ \\ 
	\hline
	\cellcolor{gray!50}POND16 & $2 \times 10^{-1}$ & $2 \times 10^{-1}$ & $2 \times 10^{-1}$ & $2 \times 10^{-1}$ & $-$ \\ 
   \hline
\end{tabular}
}
\caption*{Table of \textit{p}-values for pairwise KDE testing on the Ponderosa individuals comparing different feature spaces.  The lower diagonal corresponds to scale factors associated with strictly symmetric branching $(\beta, \gamma)$, while the upper diagonal corresponds to scale factors associated with hydraulics and asymmetric radial branching $(\bar{\beta}, \Delta \beta)$.  Significant classification is defined as occurring when \textit{p}$ < 0.05$.  Note that no single pairwise comparison results in differentiation between Ponderosa individuals as per the $p < 0.01$ threshold.}
\end{table}

\captionsetup[table]{labelfont={bf}, labelformat={default}, labelsep=space, name={Table S\hspace{-1mm}}}
\begin{table}
\caption[Table of \textit{p}-values for KDE test on human head and torso individuals.]
{\label{tab:hht_pval}\textbf{$|$ Results of pairwise classification of the human head and torso individuals using scale factors that are related to hydraulics and fluid transport and that are associated with strictly symmetric branching ($\beta, \gamma$) or asymmetric branching ($\bar{\beta}, \Delta \beta$).}}
\resizebox{\textwidth}{!}{
 \begin{tabular}{|l|c|c|c|c|c|c|c|c|c|c|c|c|c|c|c|c|c|c|}
  \hline
  \rowcolor{gray!50}
  \backslashbox{SYM}{ASYM} & HHT01 & HHT02 & HHT03 & HHT04 & HHT05 & HHT06 & HHT07 & HHT08 & HHT09 & HHT10 & HHT11 & HHT12 & HHT13 & HHT14 & HHT15 & HHT16 & HHT17 & HHT18 \\ 
  \hline
	\cellcolor{gray!50}HHT01 & $-$ & \cellcolor{gray!25} $6 \times 10^{-1}$ & \cellcolor{gray!25} $7 \times 10^{-1}$ & \cellcolor{gray!25} $6 \times 10^{-1}$ & \cellcolor{gray!25} $4 \times 10^{-1}$ & \cellcolor{gray!25} $4 \times 10^{-1}$ & \cellcolor{gray!25} $3 \times 10^{-2}$ & \cellcolor{gray!25} $5 \times 10^{-1}$ & \cellcolor{gray!25} $6 \times 10^{-1}$ & \cellcolor{gray!25} $5 \times 10^{-1}$ & \cellcolor{gray!25} $1 \times 10^{-1}$ & \cellcolor{gray!25} $5 \times 10^{-1}$ & \cellcolor{gray!25} $7 \times 10^{-1}$ & \cellcolor{gray!25} $3 \times 10^{-1}$ & \cellcolor{gray!25} $2 \times 10^{-1}$ & \cellcolor{gray!25} $6 \times 10^{-1}$ & \cellcolor{gray!25} $6 \times 10^{-1}$ & \cellcolor{gray!25} $6 \times 10^{-1}$ \\ 
	\hline
	\cellcolor{gray!50}HHT02 & $3 \times 10^{-1}$ & $-$ & \cellcolor{gray!25} $7 \times 10^{-1}$ & \cellcolor{gray!25} $7 \times 10^{-1}$ & \cellcolor{gray!25} $7 \times 10^{-1}$ & \cellcolor{gray!25} $3 \times 10^{-1}$ & \cellcolor{gray!25} $6 \times 10^{-2}$ & \cellcolor{gray!25} $7 \times 10^{-1}$ & \cellcolor{gray!25} $5 \times 10^{-1}$ & \cellcolor{gray!25} $5 \times 10^{-1}$ & \cellcolor{gray!25} $5 \times 10^{-2}$ & \cellcolor{gray!25} $5 \times 10^{-1}$ & \cellcolor{gray!25} $6 \times 10^{-1}$ & \cellcolor{gray!25} $2 \times 10^{-1}$ & \cellcolor{gray!25} $2 \times 10^{-1}$ & \cellcolor{gray!25} $6 \times 10^{-1}$ & \cellcolor{gray!25} $7 \times 10^{-1}$ & \cellcolor{gray!25} $7 \times 10^{-1}$ \\ 
	\hline
	\cellcolor{gray!50}HHT03 & $5 \times 10^{-1}$ & $3 \times 10^{-1}$ & $-$ & \cellcolor{gray!25} $8 \times 10^{-1}$ & \cellcolor{gray!25} $6 \times 10^{-1}$ & \cellcolor{gray!25} $3 \times 10^{-1}$ & \cellcolor{gray!25} $5 \times 10^{-2}$ & \cellcolor{gray!25} $7 \times 10^{-1}$ & \cellcolor{gray!25} $7 \times 10^{-1}$ & \cellcolor{gray!25} $7 \times 10^{-1}$ & \cellcolor{gray!25} $1 \times 10^{-1}$ & \cellcolor{gray!25} $6 \times 10^{-1}$ & \cellcolor{gray!25} $7 \times 10^{-1}$ & \cellcolor{gray!25} $2 \times 10^{-1}$ & \cellcolor{gray!25} $4 \times 10^{-1}$ & \cellcolor{gray!25} $7 \times 10^{-1}$ & \cellcolor{gray!25} $7 \times 10^{-1}$ & \cellcolor{gray!25} $7 \times 10^{-1}$ \\ 
	\hline
	\cellcolor{gray!50}HHT04 & $6 \times 10^{-1}$ & $4 \times 10^{-1}$ & $4 \times 10^{-1}$ & $-$ & \cellcolor{gray!25} $6 \times 10^{-1}$ & \cellcolor{gray!25} $4 \times 10^{-1}$ & \cellcolor{gray!25} $3 \times 10^{-2}$ & \cellcolor{gray!25} $7 \times 10^{-1}$ & \cellcolor{gray!25} $6 \times 10^{-1}$ & \cellcolor{gray!25} $5 \times 10^{-1}$ & \cellcolor{gray!25} $5 \times 10^{-2}$ & \cellcolor{gray!25} $6 \times 10^{-1}$ & \cellcolor{gray!25} $5 \times 10^{-1}$ & \cellcolor{gray!25} $2 \times 10^{-1}$ & \cellcolor{gray!25} $3 \times 10^{-1}$ & \cellcolor{gray!25} $6 \times 10^{-1}$ & \cellcolor{gray!25} $7 \times 10^{-1}$ & \cellcolor{gray!25} $7 \times 10^{-1}$ \\ 
	\hline
	\cellcolor{gray!50}HHT05 & $2 \times 10^{-1}$ & $6 \times 10^{-1}$ & $3 \times 10^{-1}$ & $4 \times 10^{-1}$ & $-$ & \cellcolor{gray!25} $2 \times 10^{-1}$ & \cellcolor{gray!25} $9 \times 10^{-3}$ & \cellcolor{gray!25} $6 \times 10^{-1}$ & \cellcolor{gray!25} $4 \times 10^{-1}$ & \cellcolor{gray!25} $4 \times 10^{-1}$ & \cellcolor{gray!25} $7 \times 10^{-3}$ & \cellcolor{gray!25} $2 \times 10^{-1}$ & \cellcolor{gray!25} $4 \times 10^{-1}$ & \cellcolor{gray!25} $1 \times 10^{-1}$ & \cellcolor{gray!25} $3 \times 10^{-1}$ & \cellcolor{gray!25} $6 \times 10^{-1}$ & \cellcolor{gray!25} $6 \times 10^{-1}$ & \cellcolor{gray!25} $7 \times 10^{-1}$ \\ 
	\hline
	\cellcolor{gray!50}HHT06 & $7 \times 10^{-2}$ & $3 \times 10^{-2}$ & $1 \times 10^{-1}$ & $6 \times 10^{-2}$ & $2 \times 10^{-2}$ & $-$ & \cellcolor{gray!25} $5 \times 10^{-1}$ & \cellcolor{gray!25} $3 \times 10^{-1}$ & \cellcolor{gray!25} $4 \times 10^{-1}$ & \cellcolor{gray!25} $3 \times 10^{-1}$ & \cellcolor{gray!25} $6 \times 10^{-1}$ & \cellcolor{gray!25} $9 \times 10^{-1}$ & \cellcolor{gray!25} $2 \times 10^{-1}$ & \cellcolor{gray!25} $6 \times 10^{-1}$ & \cellcolor{gray!25} $2 \times 10^{-1}$ & \cellcolor{gray!25} $2 \times 10^{-1}$ & \cellcolor{gray!25} $6 \times 10^{-1}$ & \cellcolor{gray!25} $3 \times 10^{-1}$ \\ 
	\hline
	\cellcolor{gray!50}HHT07 & $3 \times 10^{-1}$ & $2 \times 10^{-2}$ & $2 \times 10^{-1}$ & $2 \times 10^{-1}$ & $1 \times 10^{-2}$ & $2 \times 10^{-1}$ & $-$ & \cellcolor{gray!25} $1 \times 10^{-2}$ & \cellcolor{gray!25} $2 \times 10^{-1}$ & \cellcolor{gray!25} $3 \times 10^{-1}$ & \cellcolor{gray!25} $6 \times 10^{-1}$ & \cellcolor{gray!25} $2 \times 10^{-1}$ & \cellcolor{gray!25} $3 \times 10^{-3}$ & \cellcolor{gray!25} $2 \times 10^{-1}$ & \cellcolor{gray!25} $3 \times 10^{-1}$ & \cellcolor{gray!25} $1 \times 10^{-2}$ & \cellcolor{gray!25} $2 \times 10^{-1}$ & \cellcolor{gray!25} $7 \times 10^{-2}$ \\ 
	\hline
	\cellcolor{gray!50}HHT08 & $6 \times 10^{-1}$ & $4 \times 10^{-1}$ & $4 \times 10^{-1}$ & $6 \times 10^{-1}$ & $4 \times 10^{-1}$ & $3 \times 10^{-2}$ & $2 \times 10^{-1}$ & $-$ & \cellcolor{gray!25} $3 \times 10^{-1}$ & \cellcolor{gray!25} $5 \times 10^{-1}$ & \cellcolor{gray!25} $5 \times 10^{-2}$ & \cellcolor{gray!25} $6 \times 10^{-1}$ & \cellcolor{gray!25} $4 \times 10^{-1}$ & \cellcolor{gray!25} $3 \times 10^{-1}$ & \cellcolor{gray!25} $1 \times 10^{-1}$ & \cellcolor{gray!25} $5 \times 10^{-1}$ & \cellcolor{gray!25} $8 \times 10^{-1}$ & \cellcolor{gray!25} $8 \times 10^{-1}$ \\ 
	\hline
	\cellcolor{gray!50}HHT09 & $6 \times 10^{-1}$ & $4 \times 10^{-1}$ & $5 \times 10^{-1}$ & $6 \times 10^{-1}$ & $3 \times 10^{-1}$ & $2 \times 10^{-1}$ & $4 \times 10^{-1}$ & $5 \times 10^{-1}$ & $-$ & \cellcolor{gray!25} $3 \times 10^{-1}$ & \cellcolor{gray!25} $4 \times 10^{-1}$ & \cellcolor{gray!25} $7 \times 10^{-1}$ & \cellcolor{gray!25} $5 \times 10^{-1}$ & \cellcolor{gray!25} $5 \times 10^{-1}$ & \cellcolor{gray!25} $6 \times 10^{-1}$ & \cellcolor{gray!25} $5 \times 10^{-1}$ & \cellcolor{gray!25} $5 \times 10^{-1}$ & \cellcolor{gray!25} $6 \times 10^{-1}$ \\ 
	\hline
	\cellcolor{gray!50}HHT10 & $4 \times 10^{-1}$ & $3 \times 10^{-1}$ & $4 \times 10^{-1}$ & $3 \times 10^{-1}$ & $2 \times 10^{-1}$ & $2 \times 10^{-1}$ & $2 \times 10^{-1}$ & $4 \times 10^{-1}$ & $3 \times 10^{-1}$ & $-$ & \cellcolor{gray!25} $2 \times 10^{-1}$ & \cellcolor{gray!25} $6 \times 10^{-1}$ & \cellcolor{gray!25} $5 \times 10^{-1}$ & \cellcolor{gray!25} $1 \times 10^{-1}$ & \cellcolor{gray!25} $3 \times 10^{-1}$ & \cellcolor{gray!25} $5 \times 10^{-1}$ & \cellcolor{gray!25} $5 \times 10^{-1}$ & \cellcolor{gray!25} $5 \times 10^{-1}$ \\ 
	\hline
	\cellcolor{gray!50}HHT11 & $3 \times 10^{-1}$ & $3 \times 10^{-2}$ & $2 \times 10^{-1}$ & $1 \times 10^{-1}$ & $1 \times 10^{-2}$ & $2 \times 10^{-1}$ & $6 \times 10^{-1}$ & $2 \times 10^{-1}$ & $5 \times 10^{-1}$ & $2 \times 10^{-1}$ & $-$ & \cellcolor{gray!25} $6 \times 10^{-1}$ & \cellcolor{gray!25} $3 \times 10^{-2}$ & \cellcolor{gray!25} $7 \times 10^{-1}$ & \cellcolor{gray!25} $4 \times 10^{-1}$ & \cellcolor{gray!25} $5 \times 10^{-2}$ & \cellcolor{gray!25} $2 \times 10^{-1}$ & \cellcolor{gray!25} $5 \times 10^{-2}$ \\ 
	\hline
	\cellcolor{gray!50}HHT12 & $5 \times 10^{-1}$ & $3 \times 10^{-1}$ & $5 \times 10^{-1}$ & $5 \times 10^{-1}$ & $2 \times 10^{-1}$ & $1 \times 10^{-1}$ & $2 \times 10^{-1}$ & $5 \times 10^{-1}$ & $5 \times 10^{-1}$ & $6 \times 10^{-1}$ & $3 \times 10^{-1}$ & $-$ & \cellcolor{gray!25} $3 \times 10^{-1}$ & \cellcolor{gray!25} $8 \times 10^{-1}$ & \cellcolor{gray!25} $3 \times 10^{-1}$ & \cellcolor{gray!25} $4 \times 10^{-1}$ & \cellcolor{gray!25} $8 \times 10^{-1}$ & \cellcolor{gray!25} $6 \times 10^{-1}$ \\ 
	\hline
	\cellcolor{gray!50}HHT13 & $6 \times 10^{-1}$ & $5 \times 10^{-1}$ & $6 \times 10^{-1}$ & $7 \times 10^{-1}$ & $4 \times 10^{-1}$ & $4 \times 10^{-2}$ & $2 \times 10^{-1}$ & $6 \times 10^{-1}$ & $6 \times 10^{-1}$ & $3 \times 10^{-1}$ & $1 \times 10^{-1}$ & $4 \times 10^{-1}$ & $-$ & \cellcolor{gray!25} $1 \times 10^{-1}$ & \cellcolor{gray!25} $2 \times 10^{-1}$ & \cellcolor{gray!25} $6 \times 10^{-1}$ & \cellcolor{gray!25} $4 \times 10^{-1}$ & \cellcolor{gray!25} $5 \times 10^{-1}$ \\ 
	\hline
	\cellcolor{gray!50}HHT14 & $2 \times 10^{-1}$ & $2 \times 10^{-2}$ & $2 \times 10^{-1}$ & $1 \times 10^{-1}$ & $3 \times 10^{-2}$ & $1 \times 10^{-1}$ & $2 \times 10^{-1}$ & $2 \times 10^{-1}$ & $2 \times 10^{-1}$ & $6 \times 10^{-1}$ & $3 \times 10^{-1}$ & $5 \times 10^{-1}$ & $1 \times 10^{-1}$ & $-$ & \cellcolor{gray!25} $4 \times 10^{-1}$ & \cellcolor{gray!25} $2 \times 10^{-1}$ & \cellcolor{gray!25} $5 \times 10^{-1}$ & \cellcolor{gray!25} $3 \times 10^{-1}$ \\ 
	\hline
	\cellcolor{gray!50}HHT15 & $5 \times 10^{-1}$ & $2 \times 10^{-1}$ & $6 \times 10^{-1}$ & $4 \times 10^{-1}$ & $2 \times 10^{-1}$ & $6 \times 10^{-2}$ & $3 \times 10^{-1}$ & $5 \times 10^{-1}$ & $4 \times 10^{-1}$ & $3 \times 10^{-1}$ & $4 \times 10^{-1}$ & $5 \times 10^{-1}$ & $4 \times 10^{-1}$ & $3 \times 10^{-1}$ & $-$ & \cellcolor{gray!25} $2 \times 10^{-1}$ & \cellcolor{gray!25} $3 \times 10^{-1}$ & \cellcolor{gray!25} $3 \times 10^{-1}$ \\ 
	\hline
	\cellcolor{gray!50}HHT16 & $5 \times 10^{-1}$ & $6 \times 10^{-1}$ & $4 \times 10^{-1}$ & $6 \times 10^{-1}$ & $5 \times 10^{-1}$ & $2 \times 10^{-2}$ & $5 \times 10^{-2}$ & $5 \times 10^{-1}$ & $5 \times 10^{-1}$ & $3 \times 10^{-1}$ & $6 \times 10^{-2}$ & $3 \times 10^{-1}$ & $6 \times 10^{-1}$ & $6 \times 10^{-2}$ & $4 \times 10^{-1}$ & $-$ & \cellcolor{gray!25} $5 \times 10^{-1}$ & \cellcolor{gray!25} $7 \times 10^{-1}$ \\ 
	\hline
	\cellcolor{gray!50}HHT17 & $4 \times 10^{-1}$ & $6 \times 10^{-1}$ & $6 \times 10^{-1}$ & $5 \times 10^{-1}$ & $5 \times 10^{-1}$ & $1 \times 10^{-1}$ & $3 \times 10^{-1}$ & $6 \times 10^{-1}$ & $5 \times 10^{-1}$ & $3 \times 10^{-1}$ & $4 \times 10^{-1}$ & $5 \times 10^{-1}$ & $5 \times 10^{-1}$ & $2 \times 10^{-1}$ & $5 \times 10^{-1}$ & $5 \times 10^{-1}$ & $-$ & \cellcolor{gray!25} $7 \times 10^{-1}$ \\ 
	\hline
	\cellcolor{gray!50}HHT18 & $5 \times 10^{-1}$ & $3 \times 10^{-1}$ & $5 \times 10^{-1}$ & $4 \times 10^{-1}$ & $3 \times 10^{-1}$ & $9 \times 10^{-2}$ & $1 \times 10^{-1}$ & $5 \times 10^{-1}$ & $3 \times 10^{-1}$ & $7 \times 10^{-1}$ & $1 \times 10^{-1}$ & $5 \times 10^{-1}$ & $4 \times 10^{-1}$ & $4 \times 10^{-1}$ & $3 \times 10^{-1}$ & $4 \times 10^{-1}$ & $4 \times 10^{-1}$ & $-$ \\ 
   \hline
\end{tabular}
}
\caption*{Table of \textit{p}-values for pairwise KDE testing on the human head and torso individuals comparing different feature spaces.  The lower diagonal corresponds to scale factors that are related to hydraulics and space filling and that are associated with strictly symmetric branching $(\beta, \gamma)$ while the upper diagonal corresponds to scale factors associated with just hydraulics and asymmetric radial branching $(\bar{\beta}, \Delta \beta)$.  Significant classification is defined as occurring when \textit{p}$ < 0.05$.  Note that only pairwise comparisons between HHT individuals 13 and 7, 11 and 5, and 7 and 5 result in differentiation between HHT individuals as per the $p < 0.01$ threshold.  All other pairwise comparisons are indistinguishable.}
\end{table}

\captionsetup[table]{labelfont={bf}, labelformat={default}, labelsep=space, name={Table S\hspace{-1mm}}}
\begin{table}
\caption[Table of \textit{p}-values for KDE test on tree tips species.]
{\label{tab:treetips_pval}\textbf{$|$ Results of pairwise classification of tree tips species                                                                                                                                                                                using scale factors associated with strictly symmetric branching ($\beta, \gamma$), or asymmetric branching, hydraulics, and fluid transport ($\bar{\beta}, \Delta \beta$).}}
\resizebox{\textwidth}{!}{
\begin{tabular}{|l|c|c|c|c|c|c|}
  \hline
  \rowcolor{gray!50}
  \backslashbox{SYM}{ASYM} & Maple & Oak & Robinia & Whitefir & Dougfir & Whitepine \\ 
  \hline
	\cellcolor{gray!50}Maple & $-$ & \cellcolor{gray!25} $3 \times 10^{-2}$ & \cellcolor{gray!25} $7 \times 10^{-1}$ & \cellcolor{gray!25} $4 \times 10^{-10}$ & \cellcolor{gray!25} $1 \times 10^{-2}$ & \cellcolor{gray!25} $8 \times 10^{-3}$ \\ 
	\hline
	\cellcolor{gray!50}Oak & $4 \times 10^{-1}$ & $-$ & \cellcolor{gray!25} $3 \times 10^{-2}$ & \cellcolor{gray!25} $9 \times 10^{-6}$ & \cellcolor{gray!25} $4 \times 10^{-3}$ & \cellcolor{gray!25} $1 \times 10^{-1}$ \\ 
  	\hline
	\cellcolor{gray!50}Robinia & $5 \times 10^{-1}$ & $2 \times 10^{-1}$ & $-$ & \cellcolor{gray!25} $3 \times 10^{-6}$ & \cellcolor{gray!25} $4 \times 10^{-2}$ & \cellcolor{gray!25} $3 \times 10^{-2}$ \\ 
	\hline
	\cellcolor{gray!50}Whitefir & $3 \times 10^{-1}$ & $1 \times 10^{-1}$ & $4 \times 10^{-1}$ & $-$ & \cellcolor{gray!25} $2 \times 10^{-1}$ & \cellcolor{gray!25} $3 \times 10^{-1}$ \\ 
	\hline
	\cellcolor{gray!50}Dougfir & $4 \times 10^{-1}$ & $2 \times 10^{-1}$ & $5 \times 10^{-1}$ & $5 \times 10^{-1}$ & $-$ & \cellcolor{gray!25} $1 \times 10^{-1}$ \\ 
	\hline
	\cellcolor{gray!50}Whitepine & $5 \times 10^{-1}$ & $3 \times 10^{-1}$ & $6 \times 10^{-1}$ & $6 \times 10^{-1}$ & $6 \times 10^{-1}$ & $-$ \\ 
   \hline

\end{tabular}
}
\caption*{Table of \textit{p}-values for pairwise KDE testing on tree tips across species comparing different feature spaces.  The lower diagonal corresponds to scale factors that are related to hydraulics and space filling and that are associated with strictly symmetric branching $(\beta, \gamma)$ while the upper diagonal corresponds to scale factors associated with just hydraulics and asymmetric radial branching $(\bar{\beta}, \Delta \beta)$.  Significant classification is defined as occurring when \textit{p}$ < 0.05$.  All species had multiple individuals that were aggregated due to low branch number per individual.  Of note is the clustered differentiation between Angiosperm and Gymnosperm species using the radial asymmetric scale factors, specifically the Whitefir with all Angiosperms, the Dougfir with the Oak, and the Whitepine with the Maple.}
\end{table}

\clearpage

\bibliographystyle{vancouver}

\begin{thebibliography}{10}

\bibitem{price_etal_ecolett_2012}
Price CA, Weitz JS, Savage VM, Stegen J, Clarke A, Coomes DA, et~al.
\newblock Testing the metabolic theory of ecology.
\newblock Ecology letters. 2012;15(12):1465--1474.

\bibitem{alilou_etal_scireports_2018}
Alilou M, Orooji M, Beig N, Prasanna P, Rajiah P, Donatelli C, et~al.
\newblock Quantitative vessel tortuosity: A potential CT imaging biomarker for
  distinguishing lung granulomas from adenocarcinomas.
\newblock Scientific Reports. 2018;8(1):15290.

\bibitem{bentley_etal_ecolett_2013}
Bentley LP, Stegen JC, Savage VM, Smith DD, von Allmen EI, Sperry JS, et~al.
\newblock An empirical assessment of tree branching networks and implications
  for plant allometric scaling models.
\newblock Ecology Letters. 2013;16(8):1069--1078.

\bibitem{conn_etal_cellsys_2017}
Conn A, Pedmale UV, Chory J, Navlakha S.
\newblock High-Resolution Laser Scanning Reveals Plant Architectures that
  Reflect Universal Network Design Principles.
\newblock Cell Systems. 2017;5(1):53 -- 62.e3.

\bibitem{newberry_etal_ploscompbio_2015}
Newberry MG, Ennis DB, Savage VM.
\newblock Testing Foundations of Biological Scaling Theory Using Automated
  Measurements of Vascular Networks.
\newblock PLoS Computational Biology. 2015;11(8):e1004455.

\bibitem{smith_etal_newphyt_2014}
Smith DD, Sperry JS, Enquist BJ, Savage VM, McCulloh KA, Bentley LP.
\newblock Deviation from symmetrically self-similar branching in trees predicts
  altered hydraulics, mechanics, light interception and metabolic scaling.
\newblock New Phytologist. 2014;201(1):217--229.

\bibitem{savage_etal_ploscompbio_2008}
Savage VM, Deeds EJ, Fontana W.
\newblock Sizing Up Allometric Scaling Theory.
\newblock PLoS Computational Biology. 2008 09;4(9):e1000171.

\bibitem{ronellenfitsch_etal_prl_2016}
Ronellenfitsch H, Katifori E.
\newblock Global optimization, local adaptation, and the role of growth in
  distribution networks.
\newblock Physical Review Letters. 2016;117(13):138301.

\bibitem{tekin_etal_ploscompbio_2016}
Tekin E, Hunt D, Newberry MG, Savage VM.
\newblock Do Vascular Networks Branch Optimally or Randomly across Spatial
  Scales?
\newblock PLoS Computational Biology. 2016 11;12(11):1--28.

\bibitem{west_etal_science_1997}
West GB, Brown JH, Enquist BJ.
\newblock A general model for the origin of allometric scaling laws in biology.
\newblock Science. 1997;276(5309):122--126.

\bibitem{west_etal_nature_1999}
West GB, Brown JH, Enquist BJ.
\newblock A general model for the structure and allometry of plant vascular
  systems.
\newblock Nature. 1999;400(6745):664--667.

\bibitem{jain_science_2005}
Jain RK.
\newblock Normalization of tumor vasculature: an emerging concept in
  antiangiogenic therapy.
\newblock Science. 2005;307(5706):58--62.

\bibitem{prakash_etal_stroke_2013}
Prakash R, Li W, Qu Z, Johnson MA, Fagan SC, Ergul A.
\newblock Vascularization Pattern After Ischemic Stroke is Different in Control
  Versus Diabetic Rats.
\newblock Stroke. 2013;.

\bibitem{savage_etal_funceco_2004}
Savage VM, Gillooly J, Woodruff W, West G, Allen A, Enquist B, et~al.
\newblock The predominance of quarter-power scaling in biology.
\newblock Functional Ecology. 2004;18(2):257--282.

\bibitem{mori_etal_pnas_2010}
Mori S, Yamaji K, Ishida A, Prokushkin SG, Masyagina OV, Hagihara A, et~al.
\newblock Mixed-power scaling of whole-plant respiration from seedlings to
  giant trees.
\newblock Proceedings of the National Academy of Sciences.
  2010;107(4):1447--1451.

\bibitem{kolokotrones_etal_nature_2010}
Kolokotrones T, Savage V, Deeds EJ, Fontana W.
\newblock Curvature in metabolic scaling.
\newblock Nature. 2010 Apr;464(7289):753--6.

\bibitem{couinaud_liver_1957}
Couinaud C.
\newblock Le foie: {\'e}tudes anatomiques et chirurgicales.
\newblock Masson \& Cie; 1957.

\bibitem{hofmeister_plant_growth_phyllotaxy_1868}
Hofmeister WFB.
\newblock Allgemeine Morphologie der Gew{\"a}chse.
\newblock W. Engelmann; 1868.

\bibitem{metzger_etal_nature_2008}
Metzger RJ, Klein OD, Martin GR, Krasnow MA.
\newblock The branching programme of mouse lung development.
\newblock Nature. 2008;453(7196):745--750.

\bibitem{lefevre_etal_development_2017}
Lefevre JG, Short KM, Lamberton TO, Michos O, Graf D, Smyth IM, et~al.
\newblock Branching morphogenesis in the developing kidney is governed by rules
  that pattern the ureteric tree.
\newblock Development. 2017;144(23):4377--4385.

\bibitem{lau_etal_trees_2018}
Lau A, Bentley LP, Martius C, Shenkin A, Bartholomeus H, Raumonen P, et~al.
\newblock Quantifying branch architecture of tropical trees using terrestrial
  LiDAR and 3D modelling.
\newblock Trees. 2018 Oct;32(5):1219--1231.

\bibitem{newberry_angicart}
Newberry MG. Code for angicart software 2011.;.
\newblock Available from: github.com/mnewberry/angicart.

\bibitem{oliveras_etal_plantecodiv_2014}
Oliveras I, Malhi Y, Salinas N, Huaman V, Urquiaga-Flores E, Kala-Mamani J,
  et~al.
\newblock Changes in forest structure and composition after fire in tropical
  montane cloud forests near the Andean treeline.
\newblock Plant Ecology \& Diversity. 2014;7(1-2):329--340.

\bibitem{olson_etal_ecolett_2014}
Olson ME, Anfodillo T, Rosell JA, Petit G, Crivellaro A, Isnard S, et~al.
\newblock Universal hydraulics of the flowering plants: vessel diameter scales
  with stem length across angiosperm lineages, habits and climates.
\newblock Ecology Letters. 2014;17(8):988--997.

\bibitem{reich_jeco_2014}
Reich PB.
\newblock The world-wide `fast--slow'plant economics spectrum: a traits
  manifesto.
\newblock Journal of Ecology. 2014;102(2):275--301.

\bibitem{savage_etal_pnas_2010}
Savage VM, Bentley LP, Enquist BJ, Sperry JS, Smith DD, Reich PB, et~al.
\newblock Hydraulic trade-offs and space filling enable better predictions of
  vascular structure and function in plants.
\newblock Proceedings of the National Academy of Sciences. 2010
  Dec;107(52):22722--7.

\bibitem{eloy_etal_natcomm_2017}
Eloy C, Fournier M, Lacointe A, Moulia B.
\newblock Wind loads and competition for light sculpt trees into self-similar
  structures.
\newblock Nature Communications. 2017;8(1):1014.

\bibitem{brummer_etal_ploscompbio_2017}
Brummer AB, Savage VM, Enquist BJ.
\newblock A general model for metabolic scaling in self-similar asymmetric
  networks.
\newblock PLoS Computational Biology. 2017 03;13(3):1--25.

\bibitem{fluid_mechanics_landau_lifshitz_1987}
Landau L, Lifshitz E.
\newblock Fluid Mechanics. vol.~6.
\newblock 2nd ed. Pergamon, London; 1987.

\bibitem{lopez_etal_jthbio_2011}
Lopez D, Michelin S, De~Langre E.
\newblock Flow-induced pruning of branched systems and brittle reconfiguration.
\newblock Journal of Theoretical Biology. 2011;284(1):117--124.

\bibitem{moses_statistical_modeling_2017}
Moses AM.
\newblock Statistical Modeling and Machine Learning for Molecular Biology.
\newblock CRC Press; 2017.

\bibitem{friedman_elements_statistical_learning_2016}
Hastie T, Tibshirani R, Friedman J.
\newblock The Elements of Statistical Learning: Data Mining, Inference, and
  Prediction.
\newblock 2nd ed. Springer; 2016.

\bibitem{duong_jnonparstats_2013}
Duong T.
\newblock Local significant differences from nonparametric two-sample tests.
\newblock Journal of Nonparametric Statistics. 2013;25(3):635--645.

\bibitem{hunt_etal_pre_2016}
Hunt D, Savage VM.
\newblock Asymmetries arising from the space-filling nature of vascular
  networks.
\newblock Physical Review E. 2016;93(6):062305.

\bibitem{duong_etal_pnas_2012}
Duong T, Goud B, Schauer K.
\newblock Closed-form density-based framework for automatic detection of
  cellular morphology changes.
\newblock Proceedings of the National Academy of Sciences.
  2012;109(22):8382--8387.

\bibitem{warner_bullmathbio_1976}
Warner WH, Wilson TA.
\newblock Distribution of end-points of a branching network with decaying
  branch length.
\newblock Bulletin of mathematical biology. 1976;38(3):219--237.

\bibitem{banavar_etal_pnas_2010}
Banavar JR, Moses ME, Brown JH, Damuth J, Rinaldo A, Sibly RM, et~al.
\newblock A general basis for quarter-power scaling in animals.
\newblock Proceedings of the National Academy of Sciences.
  2010;107(36):15816--15820.

\bibitem{dodds_prl_2010}
Dodds PS.
\newblock Optimal form of branching supply and collection networks.
\newblock Physical Review Letters. 2010;104(4):048702.

\bibitem{barnsley_fractals_2014}
Barnsley MF.
\newblock Fractals Everywhere.
\newblock Academic press; 2012.

\bibitem{huo_etal_jrsi_2012}
Huo Y, Kassab GS.
\newblock Intraspecific scaling laws of vascular trees.
\newblock Journal of The Royal Society Interface. 2012;9(66):190--200.

\bibitem{lau_etal_forestecomanage_2019}
Lau A, Martius C, Bartholomeus H, Shenkin A, Jackson T, Malhi Y, et~al.
\newblock Estimating architecture-based metabolic scaling exponents of tropical
  trees using terrestrial LiDAR and 3D modelling.
\newblock Forest Ecology and Management. 2019;439:132--145.

\bibitem{herman_etal_plosone_2011}
Herman AB, Savage VM, West GB.
\newblock A quantitative theory of solid tumor growth, metabolic rate and
  vascularization.
\newblock PLoS One. 2011;6(9):e22973.

\bibitem{west_etal_nature_2001}
West GB, Brown JH, Enquist BJ.
\newblock A general model for ontogenetic growth.
\newblock Nature. 2001;413(6856):628.

\bibitem{barneche_etal_science_2018}
Barneche DR, Robertson DR, White CR, Marshall DJ.
\newblock Fish reproductive-energy output increases disproportionately with
  body size.
\newblock Science. 2018;360(6389):642--645.

\bibitem{mileyko_etal_plosone_2012}
Mileyko Y, Edelsbrunner H, Price CA, Weitz JS.
\newblock Hierarchical Ordering of Reticular Networks.
\newblock PLoS ONE. 2012 06;7(6):e36715.

\bibitem{katifori_etal_prl_2010}
Katifori E, Sz{\"o}ll{\H{o}}si GJ, Magnasco MO.
\newblock Damage and fluctuations induce loops in optimal transport networks.
\newblock Physical Review Letters. 2010;104(4):048704.

\bibitem{zamir_jgenphys_1978}
Zamir M.
\newblock Nonsymmetrical bifurcations in arterial branching.
\newblock The Journal of general physiology. 1978;72(6):837--845.

\bibitem{ohuchi_etal_pediares_2007}
Ohuchi H, Beighley PE, Dong Y, Zamir M, Ritman EL.
\newblock Microvascular development in porcine right and left ventricular
  walls.
\newblock Pediatric research. 2007;61(6):676--680.

\bibitem{choi_etal_physioreports_2020}
Choi JH, Kim E, Kim HY, Lee SH, Kim SM.
\newblock Allometric scaling patterns among the human coronary artery tree,
  myocardial mass, and coronary artery flow.
\newblock Physiological Reports. 2020;8(14):e14514.

\bibitem{wang_etal_physmedbio_2014}
Wang Q, Mathews AJ, Li K, Wen J, Komarov S, O'Sullivan JA, et~al.
\newblock A dedicated high-resolution {PET} imager for plant sciences.
\newblock Physics in Medicine and Biology. 2014 sep;59(19):5613--5629.

\bibitem{hubeau_etal_trendsplantsci_2015}
Hubeau M, Steppe K.
\newblock Plant-PET Scans: In Vivo Mapping of Xylem and Phloem Functioning.
\newblock Trends in Plant Science. 2015;20(10):676 -- 685.

\bibitem{moccia_etal_compmethprogbiomed_2018}
Moccia S, Momi ED, Hadji SE, Mattos LS.
\newblock Blood vessel segmentation algorithms --- Review of methods, datasets
  and evaluation metrics.
\newblock Computer Methods and Programs in Biomedicine. 2018;158:71 -- 91.

\bibitem{wang_etal_lungcancer_2017}
Wang X, Leader JK, Wang R, Wilson D, Herman J, Yuan JM, et~al.
\newblock Vasculature surrounding a nodule: a novel lung cancer biomarker.
\newblock Lung Cancer. 2017;114:38--43.

\bibitem{lambin_etal_natrevcliniconc_2017}
Lambin P, Leijenaar RT, Deist TM, Peerlings J, De~Jong EE, Van~Timmeren J,
  et~al.
\newblock Radiomics: the bridge between medical imaging and personalized
  medicine.
\newblock Nature Reviews Clinical Oncology. 2017;14(12):749.

\bibitem{pashayan_etal_science_2020}
Pashayan N, Pharoah PDP.
\newblock The challenge of early detection in cancer.
\newblock Science. 2020;368(6491):589--590.

\end{thebibliography}

\begin{thebibliography}{10}

\bibitem{s_duong_etal_pnas_2012}
Duong T, Goud B, Schauer K.
\newblock Closed-form density-based framework for automatic detection of
  cellular morphology changes.
\newblock Proceedings of the National Academy of Sciences.
  2012;109(22):8382--8387.

\bibitem{s_duong_jnonparstats_2013}
Duong T.
\newblock Local significant differences from nonparametric two-sample tests.
\newblock Journal of Nonparametric Statistics. 2013;25(3):635--645.

\bibitem{s_dray_etal_plantecol_2015}
Dray S, Josse J.
\newblock Principal component analysis with missing values: a comparative
  survey of methods.
\newblock Plant Ecology. 2015;216(5):657--667.

\bibitem{s_newberry_etal_ploscompbio_2015}
Newberry MG, Ennis DB, Savage VM.
\newblock Testing Foundations of Biological Scaling Theory Using Automated
  Measurements of Vascular Networks.
\newblock PLoS Computational Biology. 2015;11(8):e1004455.

\bibitem{s_bentley_etal_ecolett_2013}
Bentley LP, Stegen JC, Savage VM, Smith DD, von Allmen EI, Sperry JS, et~al.
\newblock An empirical assessment of tree branching networks and implications
  for plant allometric scaling models.
\newblock Ecology Letters. 2013;16(8):1069--1078.

\bibitem{s_brummer_etal_ploscompbio_2017}
Brummer AB, Savage VM, Enquist BJ.
\newblock A general model for metabolic scaling in self-similar asymmetric
  networks.
\newblock PLoS Computational Biology. 2017 03;13(3):1--25.

\bibitem{s_friedman_elements_statistical_learning_2016}
Hastie T, Tibshirani R, Friedman J.
\newblock The Elements of Statistical Learning: Data Mining, Inference, and
  Prediction.
\newblock 2nd ed. Springer; 2016.

\bibitem{s_moses_statistical_modeling_2017}
Moses AM.
\newblock Statistical Modeling and Machine Learning for Molecular Biology.
\newblock CRC Press; 2017.

\bibitem{s_tekin_etal_ploscompbio_2016}
Tekin E, Hunt D, Newberry MG, Savage VM.
\newblock Do Vascular Networks Branch Optimally or Randomly across Spatial
  Scales?
\newblock PLoS Computational Biology. 2016 11;12(11):1--28.

\bibitem{s_yang_etal_amjheartphys_2010}
Yang J, Yu L, Rennie MY, Sled JG, Henkelman RM.
\newblock Comparative structural and hemodynamic analysis of vascular trees.
\newblock American Journal of Physiology-Heart and Circulatory Physiology.
  2010;298(4):H1249--H1259.

\bibitem{s_aurelien_scikit_learn_2017}
G{\'e}ron A.
\newblock Hands-on machine learning with Scikit-Learn and TensorFlow: concepts,
  tools, and techniques to build intelligent systems.
\newblock O'Reilly Media, Inc.; 2017.

\bibitem{s_west_etal_science_1997}
West GB, Brown JH, Enquist BJ.
\newblock A general model for the origin of allometric scaling laws in biology.
\newblock Science. 1997;276(5309):122--126.

\bibitem{s_savage_etal_ploscompbio_2008}
Savage VM, Deeds EJ, Fontana W.
\newblock Sizing Up Allometric Scaling Theory.
\newblock PLoS Computational Biology. 2008 09;4(9):e1000171.

\bibitem{s_kleiber_hilgardia_1932}
Kleiber M.
\newblock Body size and metabolism.
\newblock Hilgardia. 1932 January;6(11):315--353.

\bibitem{s_savage_etal_funceco_2004}
Savage VM, Gillooly J, Woodruff W, West G, Allen A, Enquist B, et~al.
\newblock The predominance of quarter-power scaling in biology.
\newblock Functional Ecology. 2004;18(2):257--282.

\bibitem{s_mori_etal_pnas_2010}
Mori S, Yamaji K, Ishida A, Prokushkin SG, Masyagina OV, Hagihara A, et~al.
\newblock Mixed-power scaling of whole-plant respiration from seedlings to
  giant trees.
\newblock Proceedings of the National Academy of Sciences.
  2010;107(4):1447--1451.

\bibitem{s_horton_geosocbull_1945}
Horton RE.
\newblock Erosional development of streams and their drainage basins;
  hydrophysical approach to quantitative morphology.
\newblock Geological society of America bulletin. 1945;56(3):275--370.

\bibitem{s_strahler_eos_1957}
Strahler AN.
\newblock Quantitative analysis of watershed geomorphology.
\newblock Eos, Transactions American Geophysical Union. 1957;38(6):913--920.

\bibitem{s_kolokotrones_etal_nature_2010}
Kolokotrones T, Savage V, Deeds EJ, Fontana W.
\newblock Curvature in metabolic scaling.
\newblock Nature. 2010 Apr;464(7289):753--6.

\end{thebibliography}


%


%

\end{document}